\documentclass[conference]{IEEEtran}
\IEEEoverridecommandlockouts
\usepackage{cite}
\usepackage{amsmath,amssymb,amsfonts}
\usepackage{graphicx}
\usepackage{textcomp}
\usepackage{xcolor}
\usepackage{booktabs}
\usepackage{xr-hyper}
\usepackage[final]{hyperref}
\usepackage[capitalize]{cleveref}
\usepackage{algorithm}
\usepackage{algorithmicx} 
\usepackage{algpseudocode}
\usepackage[most]{tcolorbox}
\usepackage{tablefootnote}
\usepackage{booktabs}
\usepackage{array}
\usepackage{url}
\usepackage[T1]{fontenc}
\usepackage{tikz}
\usepackage{tabularray}
\usepackage{enumitem}
\UseTblrLibrary{diagbox}
\usetikzlibrary{decorations.pathmorphing,calc}

\def\BibTeX{{\rm B\kern-.05em{\sc i\kern-.025em b}\kern-.08em
    T\kern-.1667em\lower.7ex\hbox{E}\kern-.125emX}}

\newcommand{\jose}[1]{\textcolor{olive}{[jose: {#1}]}}

\NewDocumentCommand{\ARef}{ s s m }{%
    \IfBooleanTF{#2}{}{%
        \cref{#3}%
    }%
    \IfBooleanT{#1}{%
        \IfBooleanF{#2}{%
            , %
        }%
        line~\ref{#3}%
    }%
}

\begin{document}

\title{Threat Modeling for AI: The Case for an Asset-Centric Approach}

\author{\IEEEauthorblockN{Jose Rodrigo Sanchez Vicarte}
\IEEEauthorblockA{\textit{Intel Security Research} \\
jose.sanchez.vicarte@intel.com}
\and
\IEEEauthorblockN{Marcin Spoczynski}
\IEEEauthorblockA{\textit{Intel Labs} \\
marcin.spoczynski@intel.com}
\and
\IEEEauthorblockN{Mostafa Elsaid}
\IEEEauthorblockA{\textit{Intel AI Cloud Security} \\
mostafa.elsaid@intel.com}
}

\maketitle



\section{Introduction}
Recent advances in AI are transforming AI's ubiquitous presence in our world from that of standalone AI-applications into deeply integrated AI-agents.
These changes have been driven by agents' increasing capability to autonomously make decisions and initiate actions, \emph{using} existing applications; whether those applications are AI-based or not.
This evolution enables unprecedented levels of AI integration, with agents now able to take actions on behalf of systems and users - including, in some cases, the powerful ability for the AI to write and execute scripts as it deems necessary~\cite{SignificantGravitasAutoGPTV042}.
With AI systems now able to autonomously execute code, interact with external systems, and operate without human oversight, traditional security approaches fall short.

This paper introduces an asset-centric methodology for threat modeling AI systems that addresses the unique security challenges posed by integrated AI agents.
Unlike existing top-down frameworks that analyze individual attacks within specific product contexts, our bottom-up approach enables defenders to systematically identify how vulnerabilities—both conventional and AI-specific—impact critical AI assets across distributed infrastructures used to develop and deploy these agents.
This methodology allows security teams to: (1) perform comprehensive analysis that communicates effectively across technical domains, (2) quantify security assumptions about third-party AI components without requiring visibility into their implementation, and (3) holistically identify AI-based vulnerabilities relevant to their specific product context.
We have developed a threat analysis engine to automate holistic vulnerability identification (3).
This approach is particularly relevant for securing agentic systems with complex autonomous capabilities.
By focusing on assets rather than attacks, our approach scales with the rapidly evolving threat landscape while accommodating increasingly complex and distributed AI development pipelines.
To facilitate security analysis, this work also identifies eight fundamental types of AI assets and maps out the range of ways in which adversaries can impact them.


\begin{figure*}
    \centering
    \includegraphics[width=0.9\linewidth]{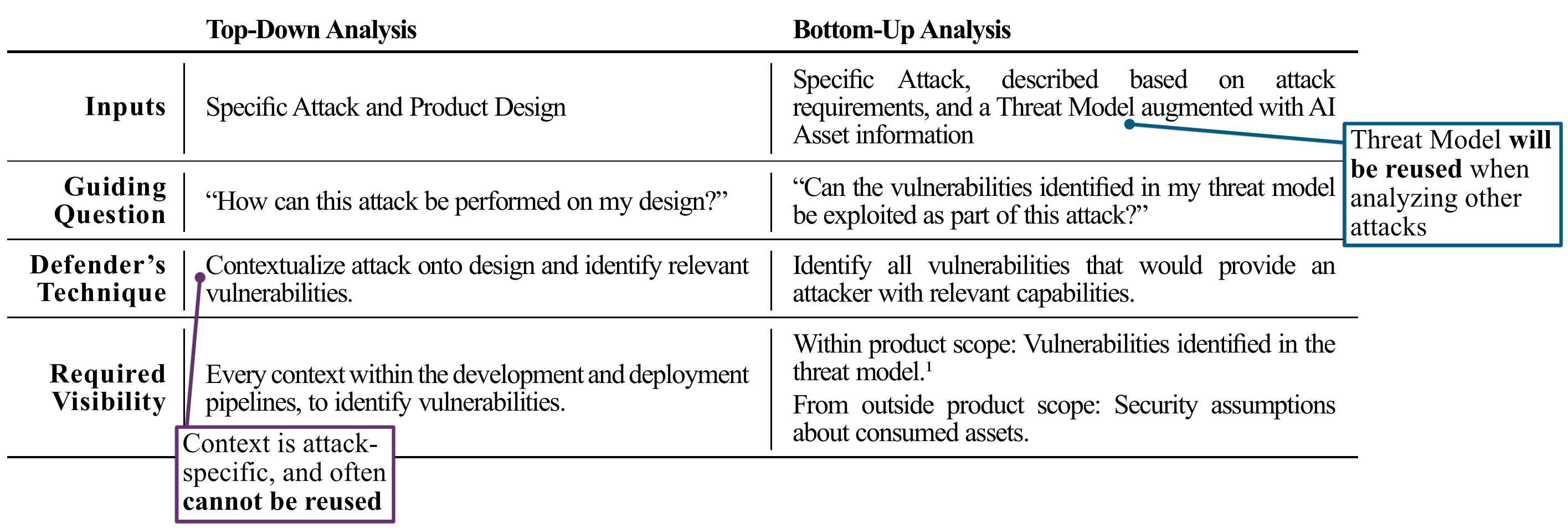}
    \caption{Compares top-down perspectives (existing frameworks) with bottom-up perspectives (our methodology) for security analysis.}
    \label{fig:td_bu_comparison}
\end{figure*}

\subsection{AI-based Vulnerabilities}
%
Integrating AI introduces vulnerabilities
into development and deployment pipelines~\cite{eulerHackingAutoGPTEscaping, OWASPTopTen, papernotSoKSecurityPrivacy2018, tabassiTaxonomyTerminologyAdversarial2019}.
These can be separated into two types:
\begin{enumerate}
    \item Conventional hardware and software vulnerabilities.
    These vulnerabilities are introduced by the hardware and software scaffolding necessary to build, run, and interact with the AI~\cite{floresgarciaRayVersions263, MaliciousAIModels}.

    \item AI-based vulnerabilities.
    These vulnerabilities occur when an adversary manipulates an AI to compromise product functionality or user data~\cite{greshakeNotWhatYouve2023}, to steal the AI itself~\cite{rakinDeepStealAdvancedModel2022, truongDataFreeModelExtraction2021}, or to steal the AI's training data~\cite{carliniExtractingTrainingData2021,choquette-chooLabelOnlyMembershipInference2021}.
\end{enumerate}

While distinct in nature, these vulnerability types are often interconnected. Conventional vulnerabilities can serve as entry points that enable AI-based attacks, while AI-based vulnerabilities can amplify the impact of conventional security breaches. For example, a conventional code injection vulnerability might allow an attacker to manipulate model inputs, enabling an AI-specific prompt injection attack which affects downstream users and agents.

Today's security processes are designed to protect development and deployment pipelines from the first type of vulnerability.
These processes must be augmented, however, to help defenders identify and reason about the second type; the ways in which adversaries can influence or manipulate an integrated AI.

Adversaries can exploit both Adversarial AI~\cite{tabassiTaxonomyTerminologyAdversarial2019} and conventional hardware and software techniques\cite{sanchezvicarteGameThreadsEnabling2020, yaoDeepHammerDepletingIntelligence2020, chenUnveilingSingleBitFlipAttacks2023} to compromise an AI.
It's critical that defenders consider both techniques in their security analysis.
That is, beyond considering the novel Adversarial AI techniques, defenders must reason about the ways in which conventional vulnerabilities can be exploited to influence AI.
To understand \emph{how} a vulnerability can be exploited to affect an AI, however, defenders must first understand \emph{what} components adversaries seek to influence.
The components – often termed Assets within threat models – which adversaries seek to manipulate inherently stem from AI's computing paradigms.

AI represents a shift in our computing paradigms towards data driven compute.
Under traditional compute, a developer would create a program – i.e. logic rules – to generate a desired output from unseen input data.
AI’s computing paradigm instead analyzes massive amounts of data to infer – i.e. learn or create – logic rules.
These inferred rules are then used to generate a desired output when applied to unseen input data.
The AI's underlying decision making logic is, therefore, algorithmically inferred rather than explicitly encoded~\cite{burkartSurveyExplainabilitySupervised2021, linardatosExplainableAIReview2021}.
This data driven nature is why AI's decisions are often difficult to explain, and why AI is difficult to validate.
Evasion, or jailbreak, attacks are an example of adversaries finding and exploiting corner cases or incorrectly correlated features within the AI's inferred logic; i.e. the model.
To compromise an AI's decision, however, adversaries are not constrained to direct attacks on the model~\cite{rakinBitFlipAttackCrushing2019,hongTerminalBrainDamage2019,mahmoudHarDNNFeatureMap2020,bober-irizarArchitecturalBackdoorsNeural2023}.
Adversaries are also able to introduce vulnerabilities into an AI by affecting its training data, training process or algorithm, and validation process or goals.
To augment their security policies against emerging AI-based vulnerabilities, defenders must reason about the many ways in which adversaries can affect their AI.
This work enables such analysis by identifying eight fundamental asset types and mapping out the range of ways in which adversaries can impact them.

\subsection{Threat Modeling Challenges}
Threat modeling for AI refers to the process of performing security analysis to identify vulnerabilities which could be exploited to impact the AI.
Defenders face three key challenges when threat modeling AI's development and deployment pipelines.

\textbf{Challenge 1.}
An AI’s implementation and dependencies are increasingly distributed across many disjoint, often cloud-based, infrastructure contexts.
Each with a unique hardware and software stack.
Defenders must consider vulnerabilities within each context.

\textbf{Challenge 2.}
Understanding all potential impacts of a vulnerability requires defenders to reason about how that vulnerability can be combined, i.e. chained, with other vulnerabilities.
To accomplish this, however, defenders must reason holistically.
Those other vulnerabilities may exist at different levels of the stack – and within different infrastructure contexts.

\textbf{Challenge 3.}
It is increasingly common for defenders to have little to no visibility into large portions of an AI's training or deployment infrastructures.
Creating an AI "from scratch" is an incredibly expensive process so developers will often \emph{consume} AI components – such as datasets or pre-trained models.
However, defenders currently lack a process for quantifying the risks they inherit by integrating those components.

Various frameworks have been proposed to facilitate top-down security analysis for AI.
Unfortunately, a top-down perspective is not well suited for addressing these challenges.
\subsection{The case for a bottom-up perspective}
\begin{figure}
    \centering
    \includegraphics[width=\linewidth]{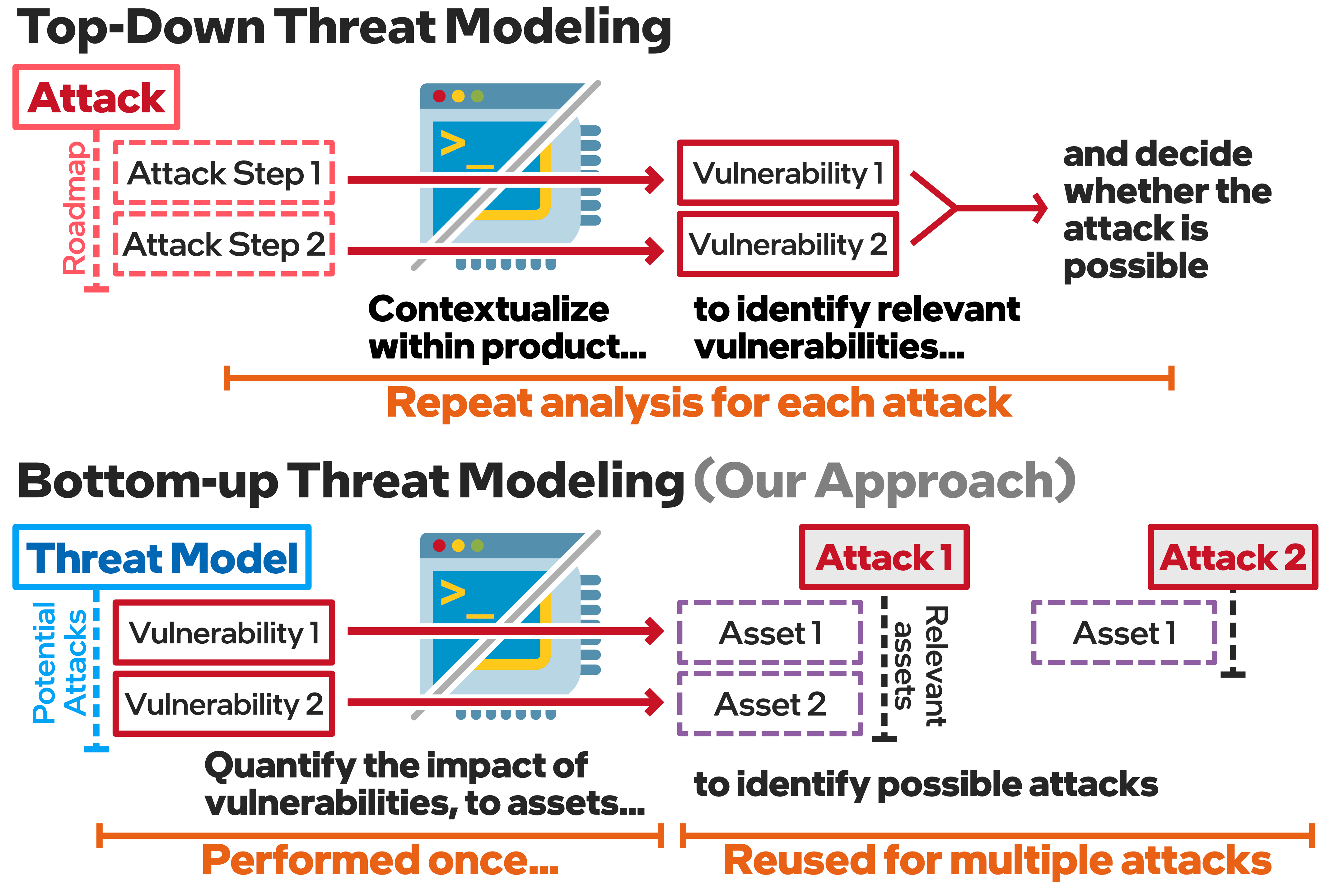}
    \caption{
    Existing frameworks enable a top-down approach (top) which allows defenders to reason about a specific attack by contextualizing it within their product.
    This work advocates for a bottom-up approach (bottom) which allows defenders to analyze their product once, from an asset-centric perspective, and reuse that analysis to contextualize \emph{multiple} attacks.
    }
    \label{fig:insight}
\end{figure}

Existing frameworks leverage a top-down perspective to individually map AI attacks within the scope of a specific product.
MITRE's ATLAS, for example, broadly describes the steps an adversary takes to perform an attack – i.e. within the attack's lifecycle – including common techniques for achieving each step.
As depicted on the top half of Figure~\ref{fig:insight}, this perspective allows defenders to reason about a specific attack by contextualizing each step in the attack's roadmap down onto their product.
The defender identifies vulnerabilities which, if exploited, would allow an adversary to perform that step.
If all attack steps can be satisfied, the attack is deemed to be in scope for the product.
To contextualize a step, however, defenders must  have enough visibility into every context of the development and deployment pipelines so as to reason about the vulnerabilities within them.
Reasoning about an attack – or even a single step – also requires reanalyzing the product.
These limitations hinder defenders from applying top-down approaches across increasingly complex AI pipelines, and prevent these approaches from scaling with a quickly growing and evolving attack space.

This work describes a methodology for \emph{bottom-up} analysis, driven by an Asset-Centric approach.
As depicted on the bottom half of Figure~\ref{fig:insight},
our methodology calls for defenders to augment their existing threat models and security analysis processes to identify \emph{how} each vulnerability can impact AI assets.
That is, defenders will identify the manner and extent – specific to the context of their product – to which an adversary can compromise those assets.
They do this for every vulnerability identified in their threat model.
Inverted from a top-down approach, defenders then use this information to identify the AI attacks which are enabled by any of the identified vulnerabilities.
Figure~\ref{fig:td_bu_comparison} summarizes the two perspectives.
This work describes \emph{how} defenders can reason about vulnerabilities' impact to AI.
Our methodology calls for defenders to:
\begin{enumerate}
    \item Augment their current security analysis to quantify whether a vulnerability – conventional or AI-based – could impact some aspect of their AI.
    This methodology is specifically designed so that defenders can easily communicate their findings across the stack and across distributed systems.

    \item Quantify, and justify, their security assumptions about the AI components consumed by their system.
    Even if the producing systems are completely closed-box.

    \item Combine the insights from their security analysis and security assumptions to identify vulnerabilities – and their root cause – relevant to the product's AI.
\end{enumerate}

Section~\ref{sec:asset_caps} introduces common types of AI Assets and maps the range of potential capabilities an adversary may obtain over each.
Section~\ref{sec:aiops} illustrates how AI Assets are built and used throughout development and deployment pipelines.
Section~\ref{sec:challenges} elaborates on the challenges defenders face when threat modeling these AI pipelines.
Section~\ref{sec:asset} provides background on asset-centric threat modeling.
Section~\ref{sec:analysis} describes our bottom-up, asset-centric, approach for threat modeling AI.
Sections~\ref{sec:enterprise_rag_bg} and~\ref{sec:case_study} introduce Enterprise RAG and apply this analysis to it, as a case study.
Section~\ref{sec:related} describes existing frameworks for threat modeling AI and why they naturally complement an asset-centric approach.
Section~\ref{sec:future} describes our ongoing work and next steps.
Finally, Section~\ref{sec:conclusion} concludes.

\section{Mapping AI Assets and Potential Capabilities}
\label{sec:asset_caps}
Defenders must adapt their security analysis to account for nuanced interactions between AI assets, stemming from AI's data-driven computing paradigm.
Under this paradigm, developers do not directly change the AI's logic rules -- i.e. the \emph{Model}.
Instead, they indirectly affect the model by changing the assets used to derive it.
It is critical, therefore, that defenders understand the different types of AI assets and the ways in which adversaries might interact with them.

This section first identifies the types of AI assets (Section~\ref{sec:asset_caps:types}), then expands conventional CIA notions into finer-grained capabilities (Section~\ref{sec:asset_caps:caps}), and finally presents various examples (Section~\ref{sec:asset_caps:examples}).

\subsection{Types of AI Assets}
\label{sec:asset_caps:types}

\textbf{This section introduces eight fundamental types of AI assets. Each type represents a critical component that adversaries may target to compromise AI systems.}

\textbf{Inputs.}
Data or information to be processed by the AI.
Inputs from benign users often require protection and isolation from other inputs.
Most adversarial AI techniques assume that attackers can submit inputs to the AI -- acting as malicious users.\\[-15pt]
\begin{tcolorbox}[colback=gray!5!white,
                  colframe=gray!100!white,
                  left=0pt,
                  right=0pt,
                  top=2pt,
                  bottom=1pt,
                  title=Common Input components include:]
\begin{itemize}[leftmargin=12pt]
    \item \textbf{User Input: The primary data or query from the user}
    \item \textbf{Prompt Context: Historical conversation or session data}
    \item \textbf{Prompt Augmentations: Additional data like RAG outputs or system prompts}
    \item \textbf{Tool/API Descriptions: Metadata about available tools and their capabilities}
    \item \textbf{Tool/API Results: Outputs from external tool calls that augment the prompt}
\end{itemize}
\end{tcolorbox}

\textbf{Outputs.}
Response(s) generated by the system.
Increasingly, outputs are consumed by another model or agent.
The output is, naturally, inherent to the functionality of the system.\\[-15pt]
\begin{tcolorbox}[colback=gray!5!white,
                  colframe=gray!100!white,
                  left=0pt,
                  right=0pt,
                  top=2pt,
                  bottom=1pt,
                  title=Common Output components include:]
\begin{itemize}[leftmargin=12pt]
    \item \textbf{Generated Text/Content: The primary AI response}
    \item \textbf{Error Messages: System feedback that may leak information}
    \item \textbf{Tool or API Actions: Requests initiated by the AI}
\end{itemize}
\end{tcolorbox}

\textbf{Output Details or Explanations.}
Additional details about the output, describing why the system provided or generated it.
Some are generated as part of inference (confidences) and others require additional evaluation (detected features).
Output details are distinguished from Outputs because adversarial techniques often require specific details or explanations.\\[-15pt]
\begin{tcolorbox}[colback=gray!5!white,
                  colframe=gray!100!white,
                  left=0pt,
                  right=0pt,
                  top=2pt,
                  bottom=1pt,
                  title=Common Output Details or Explanations components include:]
\begin{itemize}[leftmargin=12pt]
    \item \textbf{Confidence Scores: Numerical certainty measures}
    \item \textbf{Feature Attributions: Which input features influenced the output}
    \item \textbf{Decision Paths: Step-by-step reasoning chains}
\end{itemize}
\end{tcolorbox}

\textbf{Dataset.}
AI is data-driven compute, so the dataset is the most critical and sensitive asset of any AI/ML system.
Common variants include: Training, Validation, Augmentation (RAG).\\[-25pt]
\begin{tcolorbox}[colback=gray!5!white,
                  colframe=gray!100!white,
                  left=0pt,
                  right=0pt,
                  top=2pt,
                  bottom=1pt,
                  title=Datasets are composed of:]
\begin{itemize}[leftmargin=12pt]
    \item \textbf{Raw Data: Original, unprocessed information}
    \item \textbf{Encoded Data: Transformed representations (e.g., embeddings)}
    \item \textbf{Labels: Ground truth annotations for supervised learning}
    \item \textbf{Metadata: Provenance, timestamps, and quality indicators}
\end{itemize}
\end{tcolorbox}

\textbf{Model Parameters: Weights (including biases), Activations, Updates.}
The static and dynamic data values which make up the model.
Model theft, or extraction, seeks to approximate the model parameters.\\[-25pt]
\begin{tcolorbox}[colback=gray!5!white,
                  colframe=gray!100!white,
                  left=0pt,
                  right=0pt,
                  top=2pt,
                  bottom=1pt,
                  title=Key Model Parameter components include:]
\begin{itemize}[leftmargin=12pt]
    \item \textbf{Weights: Learned parameters from training}
    \item \textbf{Activations: Intermediate computational states during inference}
    \item \textbf{Updates: Gradients or parameter changes during training/fine-tuning}
\end{itemize}
\end{tcolorbox}

\textbf{Validation or Monitoring Criteria.}
Validation is a critical aspect of AI which increasingly relies on third party tools, services, and actors.
Validation involves discovery of bias and corner cases in the model and training data.
It's common practice to validate the dataset during Data Collection \& Assembly -- before incurring training costs.
Once deployed, model performance is also monitored; allowing developers to understand model performance on real-world data.\\[-25pt]
\begin{tcolorbox}[colback=gray!5!white,
                  colframe=gray!100!white,
                  left=0pt,
                  right=0pt,
                  top=2pt,
                  bottom=1pt,
                  title=Validation or Monitoring Criteria examples include:]
\begin{itemize}[leftmargin=12pt]
    \item \textbf{User Input: The primary data or query from the user}
    \item \textbf{Prompt Context: Historical conversation or session data}
    \item \textbf{Prompt Augmentations: Additional data like RAG outputs or system prompts}
    \item \textbf{Tool/API Descriptions: Metadata about available tools and their capabilities}
    \item \textbf{Tool/API Results: Outputs from external tool calls that augment the prompt}
\end{itemize}
\end{tcolorbox}

\textbf{Validation or Monitoring Results.}
Validation findings motivate additional data gathering or augmentation, changes to hyperparameters, or inform Guardrail selection.
Validation results contain highly valuable knowledge for any bad actor, as bias and corner cases can serve as invaluable guides for subsequent attacks.
Monitoring results also inform data collection and re-training.\\[-18pt]
\begin{tcolorbox}[colback=gray!5!white,
                  colframe=gray!100!white,
                  left=0pt,
                  right=0pt,
                  top=2pt,
                  bottom=1pt,
                  title=Typical validation or monitoring results include:]
\begin{itemize}[leftmargin=12pt]
    \item \textbf{Performance Reports: Accuracy, precision, recall metrics}
    \item \textbf{Error Analysis: Patterns of model failures}
    \item \textbf{Drift Detection: Changes in model behavior over time}
\end{itemize}
\end{tcolorbox}

\textbf{Hyperparameters: Model, Data, Training.}
Many hyperparameters exist throughout an AI System.
They are related to the model design, data processing, or training configuration.\\[-18pt]
\begin{tcolorbox}[colback=gray!5!white,
                  colframe=gray!100!white,
                  left=0pt,
                  right=0pt,
                  top=2pt,
                  bottom=1pt,
                  title=Hyperparameter categories include:]
\begin{itemize}[leftmargin=12pt]
    \item \textbf{Model Architecture: Layer types, sizes, connections}
    \item \textbf{Data Processing: Normalization methods, augmentation strategies}
    \item \textbf{Training Configuration: Learning rates, batch sizes, optimization algorithms}
\end{itemize}
\end{tcolorbox}

\subsection{Components}
\label{sec:asset_caps:types:components}

\textbf{Each asset type is assembled from individual components that can be reasoned about separately for security analysis.}
For example, a dataset is made up of many individual pieces of data which have been encoded, transformed, and labeled.
Common and example components are discussed throughout this work.
During security analysis, components are assets themselves.
However, the components of an asset type are ultimately a function of an AI's functionality, implementation, and use within a specific system.
To improve the generalizability and scalability of our vulnerability and attack analysis (Section~\ref{sec:analysis}), therefore, our approach relies on the affected asset types and not on their individual components.
This allows defenders to reason about the components of each affected asset type
based on their unique presentations within their system.

\textbf{Key Principle: Capabilities over components translate to (typically weaker) capabilities over the whole asset.}
Gaining a capability over a component allows an adversary to gain a capability over the asset as a whole -- however, the capability obtained over the whole asset is typically weaker than that gained over the individual component.
Prompt injection attacks epitomize this concept.
MCP Exploits, for example, are enabled by an adversary's ability to \emph{Make Arbitrary Changes} to MCP tool descriptions,
which are used as an Input component.
This provides the adversary with the ability to \emph{Contribute} to the Input~\cite{pearcyExploitingMCPTool2025}.
That is, the adversary lacks the ability to directly read or write to the Input as a whole, but is able to add data to it through tool descriptions.
\textbf{An adversary may ultimately need to compromise multiple components to obtain a desired capability over an asset type.}

\subsection{Dependencies}
\label{sec:asset_caps:types:dependencies}

\textbf{Asset types exhibit dependencies that allow capabilities to propagate across the AI lifecycle.}
Obtaining a capability over an asset's dependency allows an adversary to impact the downstream, dependent, asset -- albeit indirectly and typically with a reduced capability.

\textbf{Common dependency chains include:}
\begin{itemize}
    \item \textbf{Training Dataset → Model:} The ability to \emph{make arbitrary changes} to the training dataset allows an adversary to \emph{influence} the trained model
    \item \textbf{Input → Output:} The ability to \emph{make arbitrary changes} to an inference input allows an adversary to \emph{influence} the output
    \item \textbf{Validation Results → Hyperparameters:} Manipulated validation results can \emph{influence} hyperparameter selection
    \item \textbf{Model → Output Details:} A compromised model affects all explanations and confidence scores
\end{itemize}

\textbf{This dependency analysis enables modular security reasoning:} defenders can focus on assets within their scope while understanding how upstream compromises might affect them.
A defender's security analysis can focus solely on the asset types that exist within a specific scope, or context, because downstream defenders can translate their analysis; quantifying the impact to dependent assets within downstream scopes or contexts.
This greatly facilitates security analysis as dependencies or dependent assets may not even exist within a defender's limited scope.
Furthermore, identifying impacted dependencies allows defenders to reason about an attack's root cause.

\subsection{Relationships}
\label{sec:asset_caps:types:relationships}

\textbf{Asset types also exhibit relationships across models or agents in compound AI systems.}
The simplest of these is when an Output is directly used as an Input to another system; for example, when generated content is consumed by a Guardrail.
To accurately quantify adversarial capabilities, defenders must be careful to identify transitions across models and agents.

\textbf{Common cross-agent relationships:}
\begin{itemize}
    \item \textbf{Embedding Model → LLM:} The ability to \emph{make arbitrary changes} to a prompt before it's processed by an embedding model provides the ability to \emph{influence} the embedding model's output. Because the embedding model's output is the LLM's input, the adversary can \emph{influence} the LLM input.
    \item \textbf{LLM → Guardrail:} LLM outputs directly become guardrail inputs, transferring capabilities
    \item \textbf{RAG → LLM:} RAG outputs augment user prompts, providing \emph{contribute} capability to the combined input
    \item \textbf{Agent → Agent:} In multi-agent systems, one agent's output becomes another's context
\end{itemize}

\textbf{Section~\ref{sec:aiops} builds on these definitions to describe the different ways in which assets are built and used throughout development and deployment pipelines.}


To enable security analysis, Section~\ref{sec:asset_caps:caps} expands conventional notions of Confidentiality, Integrity, and Availability (CIA) into a finer-grained range or potential capabilities.

\subsection{Potential Capabilities}
\label{sec:asset_caps:caps}
To quantify the nuanced ways in which adversaries can manipulate AI Assets,
this work expands the conventional notions of Confidentiality and Integrity.
We identify nine potential capabilities.
Both Confidentiality and Integrity are expanded into four capabilities.
Availability is not expanded.

\textbf{Confidentiality.}
We expand Confidentiality into four distinct capabilities which refer to an adversary's ability to \emph{read} an asset.
\begin{enumerate}
    \item \textbf{Inspect (Unlimited Read).}
    This capability is akin to the conventional definition of Confidentiality, where an adversary gains the ability read the entire asset.

    \item \textbf{Partially Inspect (Limited Read).}
    Refers to an adversary's ability to read a subset of an asset.

    \item \textbf{Indirectly Inspect (Indirect or Inferred Read).}
    Refers to an adversary's ability to infer properties of an asset based on its usage or some other observable outcome.
    Adversaries may be able to reconstruct parts of the asset from this information -- i.e. obtaining the ability to \emph{partially inspect} it.
    Side channels often provide this capability.

    \item \textbf{Monitor.}
    Refers to an adversary's ability to observe whether an asset is consumed, and how often.
    The adversary, however, may lack the ability to directly or indirectly inspect the asset.
\end{enumerate}

\textbf{Integrity.}
We expand Integrity into four distinct capabilities which refer to an adversary's ability to \emph{write} to an asset.
\begin{enumerate}
    \item \textbf{Make Arbitrary Changes (Unlimited Write).}
    This capability is akin to the conventional definition of Integrity, where an adversary gains the ability to modify the entire asset.
    That is, the adversary can change any part of the asset, without constraints.
    Adversary's may self-limit to ensure that the asset retains functionality; e.g. remains the same data type.

    \item \textbf{Make Limited Changes (Limited Write).}
    Refers either to an adversary's ability to change a subset of an asset or to constraints on their changes; often, to both.
    Even with the ability to make arbitrary changes, adversaries must often ensure that any modifications are imperceptible or hidden~\cite{turnerCleanLabelBackdoorAttacks,sanchezvicarteDoubleCrossAttacksSubverting2021}.

    \item \textbf{Influence (Indirect Write).}
    Refers to an adversary's ability to indirectly affect an asset.
    This capability is typically obtained by affecting a related asset.
    An adversary may also obtain this capability by exploiting fault injection vulnerabilities.

    \item \textbf{Contribute (Write without Read).}
    This capability is uniquely prevalent, and devastating, for AI.
    It refers to an adversary's ability to add data into an asset without being able to read the asset.
    This is the capability which enables prompt injections and some poisoning~\cite{shanNightshadePromptSpecificPoisoning2024} or backdoor~\cite{sanchezvicarteDoubleCrossAttacksSubverting2021} attacks.
\end{enumerate}

\textbf{Availability.}
We do not expand the notion of availability beyond its conventional definition.
\begin{enumerate}
    \item \textbf{Availability (Withhold).}
    Refers to an adversary's ability to prevent access to an asset.
    Often achieved through a denial of service or by impacting integrity; e.g. deleting an asset.
    Throughout this work, we use \emph{withhold} when referring to adversarial impact on availability.

\end{enumerate}

As with conventional CIA, adversaries can also leverage one capability to obtain another.
For example, consider a model fingerprinting attack~\cite{yanCacheTelepathyLeveraging2020}.
First, an adversary leverages a timing channel to profile the model's execution time; \emph{indirectly inspecting} side effects of the model architecture.
The adversary then correlates timing information to infer the model architecture; obtaining the ability to \emph{inspect} it.
Critically, the second step is necessary for the \emph{indirectly inspected} information to become useful.
Constant time programming techniques, for example, mitigate leakage through timing side channels by ensuring that no such correlation exists.

\subsubsection{Implied Capabilities, for Vulnerabilities}
\label{sec:asset_caps:caps:implied}
\begin{figure}
    \centering
    \includegraphics[width=\linewidth]{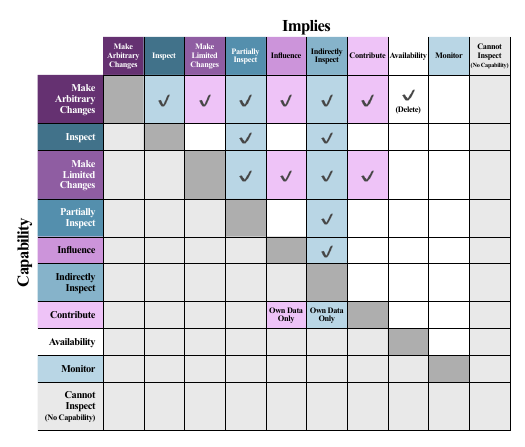}
    \caption{
    Obtaining certain capabilities (rows) provides adversaries with weaker, related, capabilities (columns).
    Capabilities are ordered from strongest (top-left) to weakest (bottom-right).
    Capabilities related to confidentiality are denoted in blues while those related to integrity are denoted in purple.
    A lack of capabilities -- i.e. \emph{cannot inspect} -- is also included for clarity.
    }
    \label{fig:implied_caps}
\end{figure}

To facilitate analysis, Figure~\ref{fig:implied_caps} also defines which capabilities are implicitly granted alongside the capability provided by a vulnerability.
That is, if a vulnerability provides a strong capability, it also provides weaker ones.
Notably, these definitions are designed to be conservative.
This allows defenders to decide whether a specific capability is out of scope instead of analyzing each weaker capability to determine whether they are in scope.

The most straightforward examples are the ability to \emph{make limited changes} by possessing the ability to \emph{make arbitrary changes} or to \emph{partially inspect} by possessing the ability to \emph{inspect}.
More subtle is an implied ability to \emph{influence} or to \emph{contribute}.
It's important to note that attackers can perform an attack which requires that they \emph{influence} or \emph{contribute} to an asset because they possess a stronger capability.
As the attack surface providing the stronger capability is likely different, attackers will likely need to adapt their technique for that attack point and stronger capability.

Note that the notion of implied capabilities only exists for vulnerabilities.
When describing how an adversary must compromise an asset, for an attack, the minimum capability is used.
This is particularly relevant to the knowledge base discussed in Section~\ref{sec:analysis:s3_attack}.
It does not make sense to reason about weaker capabilities when analyzing an attack because those capabilities would not enable the attack.
Intuitively, this property is often seen in the offensive security literature: attack variants which require weaker capabilities -- i.e. threat models or assumptions -- are published as novel attacks.
However, obtaining a stronger capability than required does enable the attack.

\subsection{Examples: Applying Capabilities to Real Scenarios}
\label{sec:asset_caps:examples}

\textbf{This section demonstrates how the capability model applies to real-world AI security scenarios, showing how capabilities over components translate to capabilities over complete assets.}

\subsubsection{Example 1: Dataset Poisoning Through Multiple Sources}
\label{sec:asset_caps:example:dataset}

\textbf{Consider a dataset $D$ composed of data from two sources: $D_1$ (adversary-controlled) and $D_2$ (trusted source).}

\textbf{Scenario:} An adversary controls the data source for $D_1$ but their malicious modifications must pass manual inspection during labeling~\cite{sanchezvicarteDoubleCrossAttacksSubverting2021}.

\textbf{Capability Analysis:}
\begin{itemize}
    \item \textbf{Over $D_1$:} The adversary has \emph{Make Limited Changes} because modifications must remain imperceptible to pass inspection
    \item \textbf{Over $D_2$:} The adversary has \emph{Cannot Inspect} - no access to this trusted source
    \item \textbf{Over complete dataset $D$:} The adversary has \emph{Contribute} - they can add malicious data through $D_1$ without seeing $D_2$
\end{itemize}

\textbf{Security Insight:} This example demonstrates how capabilities over components translate to capabilities over the whole asset. Even without full dataset access, an adversary can still poison the training process.

\subsubsection{Example 2: Prompt Injection via Tool Descriptions}
\label{sec:asset_caps:example:prompt}

\textbf{Consider a User Prompt input composed of multiple components in an LLM with tool access.}

\textbf{Scenario:} An adversary exploits a vulnerability in how tool descriptions are processed~\cite{pearcyExploitingMCPTool2025}.

\textbf{Capability Analysis:}
\begin{itemize}
    \item \textbf{Over tool descriptions:} Adversary has \emph{Make Arbitrary Changes}
    \item \textbf{Over user's original prompt:} Adversary has \emph{Cannot Inspect}
    \item \textbf{Over complete prompt (after augmentation):} Adversary has \emph{Contribute}
\end{itemize}

\textbf{Attack Path:} The adversary injects malicious instructions through tool descriptions, which get concatenated with the user's legitimate prompt, effectively hijacking the LLM's behavior without seeing the original query.

\subsubsection{Example 3: Model Extraction Through API Access}
\label{sec:asset_caps:example:extraction}

\textbf{Consider an LLM deployment where users can query the model through an API.}

\textbf{Scenario:} An adversary seeks to steal the model by observing input-output pairs.

\textbf{Capability Analysis:}
\begin{itemize}
    \item \textbf{Over Model Parameters:} Adversary has \emph{Indirectly Inspect} through API responses
    \item \textbf{Over Inputs:} Adversary has \emph{Make Arbitrary Changes} (can craft any query)
    \item \textbf{Over Outputs:} Adversary has \emph{Inspect} (sees all responses to their queries)
    \item \textbf{Over Output Details:} Depends on API configuration - may have access to confidence scores
\end{itemize}

\textbf{Security Insight:} The combination of arbitrary input control and output observation enables model extraction attacks, even without direct parameter access.

\textbf{Key Takeaway:} Capabilities degrade but propagate across model boundaries. Understanding these transitions is crucial for comprehensive security analysis.

\section{Processing and Using AI Assets}\label{sec:aiops}
This section provides a comprehensive analysis of how assets and their associated capabilities evolve through the AI lifecycle.
AI assets take different forms as they are built and used across AIOps stages -- affecting the capabilities adversaries can obtain.
This section identifies the unique set of assets, and potential capabilities, which exist across three representative AIOps stages:
\begin{enumerate}
    \item Dataset Collection \& Assembly. (Section~\ref{sec:aiops:coll&asm})
    \item Training. (Section~\ref{sec:aiops:train})
    \item Deploy \& Inference. (Section~\ref{sec:aiops:deploy})
\end{enumerate}

\textbf{Key Insight:} The same asset may offer different adversarial capabilities depending on the stage and process accessing it. Understanding these variations is crucial for comprehensive threat modeling.
Common processes and operations are also identified for each illustrative stage.
In practice, it is not common for all stages, processes, or even operations to be implemented as a single monolithic block~\cite{mcgrawARCHITECTURALRISKANALYSIS}.
Pipelines also often exhibit
circular transitions between processes and stages.
For example, within the Training stage, the Training and Model Analysis processes will interleave many times before the AI is deemed ready to be deployed.
Model Analysis could also reveal a need for additional data – requiring an additional iteration through a Dataset Collection \& Assembly stage.
It's also common to encounter unique variants of a stage, process, or operation.
For example, developers must add their own variants of the Dataset Collection \& Assembly and Training stages to fine-tuning a pre-trained model.

Due to space constraints,
this section performs a deep dive into Dataset Collection \& Assembly and provides a high level overview of both Training and Deploy \& Inference.

\subsection{Dataset Collection \& Assembly}
\label{sec:aiops:coll&asm}
\begin{figure}
    \centering
    \includegraphics[width=\linewidth]{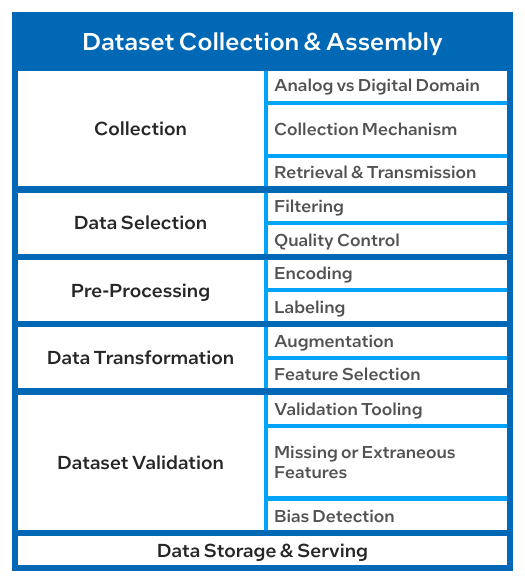}
    \caption{
    Common processes (left column) and
    operations or components (right column)
    used for Dataset Collection \& Assembly.
    It can be useful to consider 'Validation' as a separate stage when analyzing a specific pipeline's implementation, because it is often implemented on separate infrastructure or even outsourced to validation teams and services.
    }
    \label{fig:coll_asm}
\end{figure}

Dataset Collection \& Assembly encompasses the processes for collecting \emph{Raw Data} and assembling it into a \emph{Dataset}.
Processes and operations for this stage are summarized in Figure~\ref{fig:coll_asm}. We distinguish between \emph{processes} (high-level activities like "Data Selection") and \emph{operations} (specific implementations like "Filtering" or "Quality Control").
Assets present within this stage are:
\begin{enumerate}
    \item \textbf{Dataset:} Composed of processed Raw Data components.
    \item \textbf{Hyperparameters:} Configuration for data selection, pre-processing, transformation
    \item \textbf{Dataset Validation Criteria:} Rules and thresholds for data quality
    \item \textbf{Dataset Validation Results:} Outputs from validation processes
\end{enumerate}

The Dataset is assembled from \emph{Raw Data} components which have been processed; i.e. encoded, transformed, and labeled.
As discussed in Section~\ref{sec:asset_caps:types:components}, a capability over some raw data assets translates as a weaker capability over the dataset as a whole.
For example, \emph{making arbitrary changes} over a subset of raw data translates into \emph{making limited changes} to the dataset.
Common processes and  operations for
this stage
are outlined in Figure~\ref{fig:coll_asm}.

\textbf{Reasoning about Processes.}
Although the implementation of each process is unique to each pipeline, compromising
the Collection, Data Selection, Pre-Processing, or Data Transformation processes
typically allows an adversary to affect any property of raw data.
This is because each of these processes is inherently designed to modify the data it consumes.
Any impact to Confidentiality, Integrity, or Availability propagates when subsequent processes consume this modified data.
\begin{itemize}
    \item \textbf{Collection} consumes \textbf{raw data} from a source external to the system.
    
    Outputs \textbf{raw data}.
    
    \item \textbf{Data Selection} consumes \textbf{raw data} and \textbf{data selection hyperparameters}.
    This processes discards raw data which does not meet selection criteria.
    
    Outputs \textbf{selected data}.
    
    \item \textbf{Pre-Processing} consumes \textbf{selected data} and \textbf{pre-processing hyperparameters}.
    
    Outputs \textbf{pre-processed data}.
    
    \item \textbf{Transformation} consumes \textbf{pre-processed data} and \textbf{transformation hyperparameters}.
    
    Outputs \textbf{transformed data}.

\end{itemize}

Certain processes inherently limit the capabilities which can be obtained.
Dataset Validation, for example, loads the dataset but does not need to write it back.
Exploits affecting dataset integrity within this process, therefore, would not propagate.
However, an adversary could compromise the validation criteria or results to \emph{influence} the dataset -- exploiting a dependency chain.
This same inherent limitation typically applies to all hyperparameters consumed within this stage.
For capabilities to be permanent, instead of transient, adversaries must instead compromise these assets outside these processes.
\begin{itemize}
    \item \textbf{Dataset Validation} consumes an assembled \textbf{dataset} and \textbf{validation criteria}.
    
    Outputs \textbf{validation results}.
\end{itemize}

\textbf{Affecting processes through Operations.}
To compromise a specific process, adversaries leverage the attack surfaces exposed by the operations implementing the process.
Due to space constraints, we only provide examples for a small set of operations within Collection.
The following are example attack surfaces exposed as raw data is collected through a camera sensor on an edge device.
Collected data is buffered in local storage before being transferred to a centralized data storage server; e.g. an S3 bucket
The other processes within Dataset Collection \& Assembly are applied from this server.
Attack surfaces are grouped based on the operation which exposes them.
The specific vulnerabilities exploited at each surface would depend on that operation's implementation and may exist at any point across its hardware/software stack.
\begin{itemize}
    \item \textbf{Analog vs Digital Domain.}
    An adversary might affect data within the analog domain, before it is collected by the camera sensor.
    As collected data will undergo selection and pre-processing, adversaries are constrained to \emph{making limited changes}.

    \item \textbf{Collection Mechanism.}
    An adversary might affect the \emph{availability} of data by disabling the camera sensor.
    They might also \emph{make limited changes} to the camera's output by exploiting a vulnerability in the camera's hardware implementation or drivers.

    \item \textbf{Retrieval \& Transmission.}
    An adversary might \emph{make limited changes} to collected data while it's buffered in local storage or intercept it in transit from the data collection endpoint to the central server.
\end{itemize}

\subsection{Training}
\label{sec:aiops:train}
\begin{figure}
    \centering
    \includegraphics[width=\linewidth]{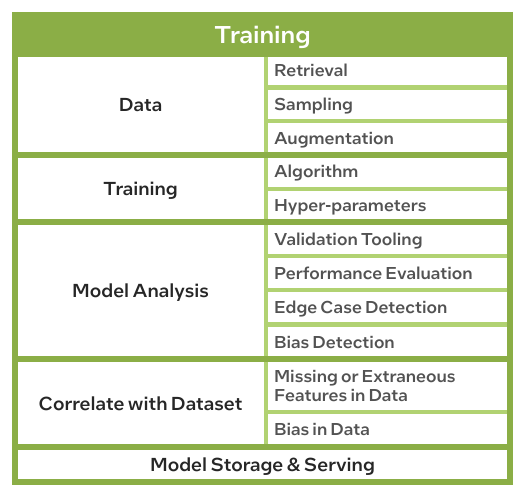}
    \caption{
    Common processes (left column) and operations or components (right column) used for Training.
    For the same reasons as within Dataset Collection \& Assembly, it is often useful to consider Model Analysis and Correlation with Dataset as a separate 'Validation' stage when analyzing a specific pipeline's implementation.
    }
    \label{fig:training}
\end{figure}

The Training stage encompasses the processes used to create or update a Model. This stage presents unique security challenges due to the iterative nature of the training process and the complex interactions between data, model, and hyperparameters.
Common operations are outlined in Figure~\ref{fig:training}.
Assets present within this stage are:
\begin{enumerate}
    \item \textbf{Datasets: Training} and \textbf{Validation.}
    \item \textbf{Inputs:} Batched and augmented training samples.
    \item \textbf{Outputs} \& \textbf{Output Details.}
    \item \textbf{Hyperparameters: Learning rates, architectures, augmentation strategies.}
    \item \textbf{Model Parameters: Weights, gradients, optimizer states.}
    \item \textbf{Model Validation Criteria:} Performance thresholds and early stopping rules
    \item \textbf{Model Validation Results:} Accuracy metrics, loss values, confusion matrices
\end{enumerate}

\textbf{Critical Security Insight:} The training process typically requires only read access to datasets. This principle of least privilege can be enforced to prevent dataset tampering during training.

\textbf{Capability Constraints During Training:}
\begin{itemize}
    \item Dataset Access: Read-only by design - training processes should never modify source datasets
    \item Model Parameters: Full write access required - the core purpose of training
    \item Transient Augmentations: Temporary modifications that don't persist to storage
\end{itemize}
As within Dataset Collection \& Assembly, the set of obtainable capabilities can be inherently constrained based on which process or operation is compromised.
Augmentation, for example, modifies samples from the dataset before they are input to the model by the Training Algorithm.
These modifications are transient -- they are not written back to the dataset.
Defenders can leverage this property to implement training processes which can access and operate on the dataset but can never modify its source;
follow principles of least privilege.
Similar principles apply to the validation dataset.


\subsection{Deploy \& Inference}
\label{sec:aiops:deploy}
\begin{figure}
    \centering
    \includegraphics[width=\linewidth]{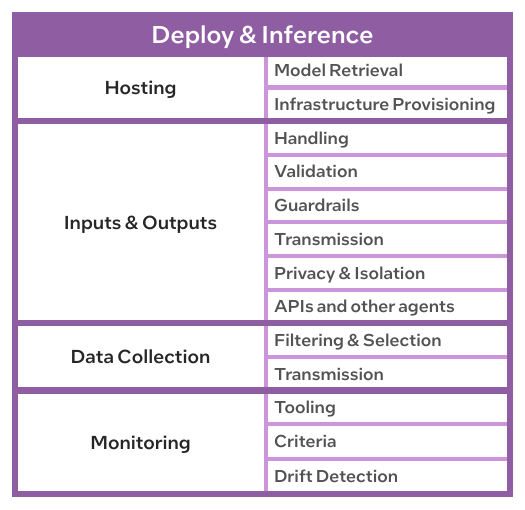}
    \caption{
    Common processes (left column) and operations or components (right column) used for deploying an AI and enabling its use.
    It is often useful to consider Data Collection and Monitoring as a separate `Monitoring' stage when analyzing a specific implementation, as it is often implemented on separate infrastructure – to facilitate aggregation – and can also be outsourced to separate teams and services.
    }
    \label{fig:deploy}
\end{figure}

This stage describes the process for deploying a model, serving inference requests, monitoring model performance, and collecting data for future re-training.
Common operations are outlined in Figure~\ref{fig:deploy}.
Assets present within this stage are:
\begin{enumerate}
    \item \textbf{Dataset: Training} (implicitly, through model).
    \item \textbf{Inputs.}
    \item \textbf{Outputs} \& \textbf{Output Details.}
    \item \textbf{Hyperparameters: Model, Data selection.}
    \item \textbf{Model Parameters: Weights, Activations.}
    \item \textbf{Monitoring Criteria.}
    \item \textbf{Monitoring Results.}
\end{enumerate}

\subsection{Summary of Capabilities Across AIOps Stages}

Table~\ref{tab:capability-summary} summarizes how adversarial capabilities vary across AIOps stages for key asset types.

\begin{table}[h]
\centering
\caption{Adversarial Capabilities by Asset Type Across AIOps Stages}
\label{tab:capability-summary}
\begin{tabular}{|l|l|l|l|}
\hline
\textbf{Asset Type} & \textbf{Collection} & \textbf{Training} & \textbf{Deployment} \\
\hline
Dataset & Full Access & Read Only & Not Present \\
Model Params & Not Present & Write Required & Read Only \\
Inputs & Write (Collection) & Read (Batches) & Write (User) \\
Outputs & Not Present & Metrics Only & Full Generation \\
\hline
\end{tabular}
\end{table}
Within Deploy \& Inference, the set of capabilities which can be obtained over the model and dataset are typically constrained.
Processes within this stage operate on the model but do not write it back, limiting how far integrity attacks can propagate.
Adversaries cannot compromise the training dataset's integrity, and cannot directly compromise its confidentiality, because it is not deployed alongside the model.
However, the training dataset is partially encoded -- memorized -- within the model weights during training.
This allows adversaries to \emph{indirectly inspect} the training dataset by studying the model's behavior~\cite{choquette-chooLabelOnlyMembershipInference2021,fredriksonModelInversionAttacks2015}.

Defenders must carefully asses the flow of assets within this stage.
For example, developers can reduce latency by moving the Data Collection process off the critical path for the Transmission process; i.e. for the process which responds to users or provides outputs to downstream systems.
This implementation decision also limits adversarial impact to integrity or availability:
Modifications within the Data Collection process would not propagate to transmitted outputs.
However, this decision does not affect impact to confidentiality.

\section{Challenges to Overcome}\label{sec:challenges}
Security analysis for AI must overcome three key challenges.
First, systems are increasingly distributed and specialized.
Second, understanding all possible impacts of a vulnerability requires a holistic evaluation.
Finally, aggravating the first two challenges, the vast majority of AI products are limited in scope but inherit risks in components consumed from outside their scope.

\textbf{
Challenge 1. AI Systems are increasingly disjoint and distributed.
}
Training famously heavily benefits from scale out compute~\cite{alistarhBriefTutorialDistributed2018, limAcceleratingTrainingDNN2017}, however, training is far from only operation implemented in a distributed fashion~\cite{mcgrawARCHITECTURALRISKANALYSIS}.
Consider Model Validation; the union of Model Analysis and Correlate with Dataset in the Training stage (Figure~\ref{fig:training}).
The complexity of validation scales alongside growing data and model sizes~\cite{Gpt4systemcardpdf,zhangUnderstandingDeepLearning2017}.
Increasingly, this complexity is addressed by specialized actors, using specialized tooling, on separate infrastructure.
Similarly, it is unlikely that the data-science experts who train the AI will be the ones who deploy and maintain its inference infrastructure.

\textbf{
Challenge 2. Understanding the impact of a vulnerability requires a holistic evaluation.
}
Distributing an AI's implementation and dependencies distributes its assets accordingly.
Consider all the processes outlined in Figures~\ref{fig:coll_asm},~\ref{fig:training}, and~\ref{fig:deploy}.
Every process can exist within a separate hardware and software context, and each performs unique operations or transformations on a subset of the system's AI Assets.
This creates many, often disjoint, attack surfaces.
Reasoning about all possible implications of a vulnerability – identified in isolation, within a single context – requires that defenders map out how that vulnerability can be chained with vulnerabilities across other contexts.


\textbf{
Challenge 3. The vast majority of AI products are limited in scope.
}
It is increasingly rare for real world products to develop an entire AI “from scratch.”
That is, for real world products to implement every stage, process, or operation necessary to train, deploy, and monitor an AI.
Instead, developers leverage existing products and services to consume components necessary for their product.
Pre-built datasets and pre-trained models epitomize this scenario.
Software libraries, including those for validation or monitoring, are another common example of consumed components.

These components and the processes used to create them are often proprietary, however, so defenders often have little to no visibility into their validation or development.
This greatly complicates defenders' ability to identify and quantify the risks they inherit by consuming those components.
Defenders currently lack a clear method for quantifying security assumptions about consumed components – limiting their ability to reason about vulnerabilities stemming from their use.

Our methodology addresses these challenges using an asset-centric approach.
Section~\ref{sec:asset} provides background on asset-centric threat modeling, and Section~\ref{sec:analysis} describes how we build on it using the fundamental types of AI assets introduced in Section~\ref{sec:asset_caps}.

\section{Asset-Centric Threat Modeling}\label{sec:asset}
Asset-Centric threat modeling presents a methodology for defenders to reason about how adversaries can impact assets and what said impact means for the system as a whole.
Asset-centric threat modeling formally defines an asset as:

\begin{tcolorbox}[colback=gray!5!white,
                  colframe=black!75!black,
                  left=0pt,
                  right=0pt,
                  top=2pt,
                  bottom=1pt,
                  title=Asset]
    A valuable object which requires protection. Defenders often distinguish between assets which are inherently valuable and those which are valuable because they protect, or serve as a stepping stone towards, other assets.
\end{tcolorbox}

Using this methodology, defenders also quantify the degree of control (Integrity or Availability) or visibility (Confidentiality) an exploit provides over an Asset.

\begin{tcolorbox}[colback=gray!5!white,
                  colframe=black!75!black,
                  left=0pt,
                  right=0pt,
                  top=2pt,
                  bottom=1pt,
                  title=Adversarial Capability over an Asset]
    Describes a specific degree of control or visibility an adversary has, or could gain, over a specific asset.
\end{tcolorbox}

Naturally, the amount of control or visibility gained depends on the vulnerability – the potential threat – being exploited.
For example, compromising a user account might allow an adversary to Read a specific file, while compromising an admin account allows them to both Read and Write to it.

\subsection{Asset-Centric Reasoning about Threats}\label{sec:asset:requirements}
Although some assets are inherently valuable, others are valuable because they protect, or serve as a stepping stone towards, other assets.
Adversaries typically compromise the latter, through individual exploits, as they move towards their true goal: compromising one or more assets which are inherently valuable.
It is therefore useful to identify the complete set of assets -- of both types -- an adversary must compromise along the way to their goal.
Including the minimum capabilities they must obtain over each asset.

Consider, as an example, theft of encrypted Personally Identifiable Information (PII) Data from storage.
The data itself is an inherently valuable asset.
The cryptographic key used to encrypt the data only has value because it protects that data.
To compromise data confidentiality, however, the adversary must obtain the capabilities listed below over both assets.\footnote{We’ll ignore exfiltration throughout this example, leaving it as an exercise for the reader to consider the additional assets – and capabilities over those assets – which the adversary must gain to enter the system and exfiltrate the data.}


\begin{itemize}
    \item Cryptographic Key (Asset): Inspect (Capability, compromising confidentiality).
    \item Encrypted PII Data (Asset): Inspect (Capability, compromising confidentiality).
\end{itemize}

In practice and in threat models, the vulnerabilities which provide each capability can be reasoned about – and exploited – as individual steps in a chain of attacks.
Adversaries might, for example, leverage side channels in a cryptographic algorithm's implementation to reconstruct the key.
Reconstructing the key is the process through which they transform the ability to \emph{indirectly inspect} it into the ability to \emph{inspect} it.
Section~\ref{sec:analysis} will describe how we leverage this property to develop our methodology.





\section{Asset-Centric Threat Modeling for AI}\label{sec:analysis}
Our asset-centric analysis, depicted in \ARef{alg:overall},
is comprised of four steps to identify relevant threats $TM$ and a fifth step to identify mitigations $M$.
This section discussions the four steps for identifying $TM$ as two phases, each comprised of two steps.

In phase one (\ARef*{alg:overall:mapAF}) defenders map $AF$ by identifying capabilities adversaries could obtain over the AI asset types relevant to their $system$.
Asset types and the set of possible capabilities are introduced in Section~\ref{sec:asset_caps}.
Defenders reason about these capabilities based on whether they originate within product scope (Section~\ref{sec:analysis:s1_vulns}) or are inherited from consumed assets (Section~\ref{sec:analysis:s2_assumptions}).
Our analysis implements this reasoning as two steps, however, there is no strict dependence between them.
They can be performed in parallel with each other and alongside existing security analysis.
The set of possible capabilities is also the same for both steps, the difference between them is whether each capability originates from a vulnerability or a security assumption.

Phase two performs threat analysis (\ARef*{alg:overall:initKB}) to generate a knowledge base of attacks $KB$, and then identifies which threats from $KB$ are relevant based on $AF$ (\ARef*{alg:overall:mapThreats}).
Section~\ref{sec:analysis:s3_attack} describes how threats are broken down into their asset-centric requirements and added into $KB$.
Section~\ref{sec:analysis:s4_mapping} describes how those threats are contextualized within a product, using $AF$, to determine if, and how, they are feasible.
We have developed a threat analysis engine that automates $MapThreats(AF, KB)$ (\ARef*{alg:overall:mapThreats}).


$KB$ is ordered first in \ARef{alg:overall} because defenders can \emph{reuse} $AF$ (\ARef*{alg:overall:mapThreats}) as $KB$ changes – i.e. as new threats are discovered.
Similarly, $KB$ can also be reused as $system$ or $AF$ change; even across different products.
These properties allow defenders to efficiently scale their analysis with the growing and evolving threat landscape.

\begin{algorithm}
\caption{Asset-Centric AI Threat Modeling: Overall Process}
\label{alg:overall}
\begin{algorithmic}[1]
\Procedure{AssetCentricModel}{$system$}
    \State $KB \gets$ InitializeKnowledgeBase() \label{alg:overall:initKB}
    \State $AF \gets$ MapAdversaryFootprint($system$) \label{alg:overall:mapAF}
    \State $TM \gets$ MapThreats($AF, KB$) \label{alg:overall:mapThreats}
    \State $M \gets$ IdentifyMitigations($TM$)  \label{alg:overall:mitigID}
    \State \Return threat analysis results
\EndProcedure
\end{algorithmic}
\end{algorithm}

Phase 1, where defenders perform security analysis to generate $AF$ for their $system$, is outlined in \ARef{alg:p1} and described in Sections~\ref{sec:analysis:s1_vulns} and~\ref{sec:analysis:s2_assumptions}.


\subsection{
Identifying Adversarial Influence within Product Scope (Phase 1)
}\label{sec:analysis:s1_vulns}
This step guides defenders through AI-centric analysis of vulnerabilities within product scope.
During this process, defenders will reason about vulnerabilities from their existing security analysis but may also identify additional related vulnerabilities, such as those described in~\cite{mcgrawARCHITECTURALRISKANALYSIS,khawajaDatabricksAISecurity2024}.
As outlined in \ARef*{alg:p1:s1_start}-\ARef**{alg:p1:s1_end}, this step guides defenders through
identifying the AI asset types that exist within each in-scope stage,
identifying vulnerabilities which could impact each asset,
quantifying the capabilities each vulnerability could provide,
and aggregating them into $AF$.

Repeating this process within each stage – i.e. each system boundary – allows $AF$ to describe capabilities which could be obtained across across the entire development and deployment pipelines.
This allows defenders to address Challenge 1 (Section~\ref{sec:challenges}).
An example output from this step is depicted on the top right of Figure~\ref{fig:flow_summary}.

\subsubsection{Holistic Evaluation}
Within \ARef*{alg:p1:determine_capabilities}, defenders may identify immediate risks based on gained capabilities.
For example, the risk of dataset theft through a vulnerability which provides the ability to read the dataset in storage.
However, identifying \emph{all} risks enabled by this vulnerability, as described in Challenge 2, requires that defenders consider the aggregated capabilities an adversary may gain across the entire product lifecycle.

\begin{algorithm}
\caption{Phase 1: Mapping Adversary Footprint}
\label{alg:p1}
\begin{algorithmic}[1]
\Procedure{MapAdversaryFootprint}{$system$} \label{alg:p1:func}
    \State $AF \gets \emptyset$
    \For{each stage $s$ in system.stages} 
        \State $inScopeAssets \gets$ IdentifyAssets($s$) \label{alg:p1:s1_start}
        \For{each asset $a$ in $inScopeAssets$}
            \State $cap \gets$ DetermineCapabilities($a, s$)
            \State Add $(a, cap)$ to $AF$
        \EndFor \label{alg:p1:s1_end}
        \State $consumedAssets \gets$ IdentifyConsumedAssets($system, s$) \label{alg:p1:s2_start}
        \If{$consumedAssets \neq \emptyset$}
            \For{each asset $a$ in $consumedAssets$}
                \State $assumptions \gets$ JustifyAssumptions($a, s$)
                \State Add $(a, assumptions)$ to $AF$ 
            \EndFor
        \EndIf \label{alg:p1:s2_end}
    \EndFor
    
    \State \Return $AF$ 
\EndProcedure

\Procedure{DetermineCapabilities}{$asset, stage$} \label{alg:p1:determine_capabilities}
    \State Identify attack surfaces related to $asset$ through existing security analysis
    \State For each attack surface, determine adversarial capabilities (Read, Write, Execute, etc.)
    \State Consider different adversary types with access to attack surface (insider, remote, etc.)
    \State \Return list of capabilities with constraints
\EndProcedure

\Procedure{JustifyAssumptions}{$asset, stage$} \label{alg:p1:justify_assumptions}
    \State Consider producer reputation, provenance information, and asset properties
    \State Document justifications for trust or distrust in the asset
    \State \Return list of capabilities with constraints
\EndProcedure
\end{algorithmic}
\end{algorithm} 

\subsubsection{Enabling Collaboration}
Threat modeling and security experts – who are not AI or AI security experts – are already applying variants of \ARef*{alg:p1:determine_capabilities} when reasoning about conventional assets.
It is, therefore, more straightforward for them to reason about AI \emph{assets}
than it is for them to identify the full range of AI \emph{threats} enabled by an exploit.
For example, a hardware security expert can identify that a fault injection attack would provide an adversary the ability to “perform transient writes to model weights, during inference."
Leveraging $KB$, they use that information to identify in scope AI threats; e.g. an evasion attack.
This information would also allow an AI security expert – who may not be a hardware security expert – to reason about the risks stemming from an exploit of that fault injection.

\subsubsection{Reasoning about Different Adversaries}
It is common practice to threat model from the perspective of multiple adversaries, each with a distinct level of access to the system or product.
In scenarios where the obtained capabilities depend on each adversary's level of access, defenders would identify each adversary-specific outcome.
Subsequent analysis, in Phase 2, would leverage this information to contextualize and distinguish each adversary's specific techniques.


\subsection{
Quantifying Assumptions about Adversarial Influence Outside Product Scope (Phase 1)
}\label{sec:analysis:s2_assumptions}
The second step in phase one, depicted in \ARef*{alg:p1:s2_start}-\ARef**{alg:p1:s2_end}, calls for defenders to identify the capabilities adversaries may have obtained over consumed assets while they were outside product scope.

At first glance, the lack of visibility into each consumed asset's source appears to invite pessimistic assumptions.
That is, defenders could argue that consuming any asset is a massive risk because they cannot guarantee that an adversary did not obtain full control over the asset outside product scope.
However, defenders \emph{are} able to justify constraints on assumed capabilities.
As described in \ARef*{alg:p1:justify_assumptions}, defenders could leverage supplemental material – such as provenance information – to justify constraints on assumed capabilities.
Effective provenance information should be end-to-end, capturing the complete lifecycle of AI assets from creation through deployment, as proposed in frameworks like Atlas~\cite{spoczynskiAtlas2024}.
Defenders could also leverage the producer’s reputation to justify constraints.
For example, defenders might justify a larger degree of trust in a producer’s infrastructure if that producer is a large AI corporation instead of an anonymous account on a public code repository or model zoo.
Similarly, properties about the asset itself may justify a larger degree of trust.
For example, it is easier to justify trust in the validation of a pre-trained model that has widespread adoption and has been the subject of academic and industry research.

The resulting assumptions are framed in the same way as if they had been identified in the previous step;
an example output is depicted on the top left of Figure~\ref{fig:flow_summary}.
This allows defenders to aggregate them with the capabilities obtained within product scope.
The top half of Figure~\ref{fig:flow_summary}, encompassing both steps in Phase 1, then depicts a complete $AF$.

\subsection{
Threat Analysis (Phase 2)
}\label{sec:analysis:s3_attack}
The first step in phase two (\ARef*{alg:overall:initKB}) generates $KB$ – a prerequisite for mapping attacks (\ARef*{alg:overall:mapThreats}).
Defenders repeat the first portion of the procedure in \ARef*{alg:p2:threat_analysis} for every attack they wish to add into $KB$.
Fortunately, this process is product-independent.
$KB$ can be reused when analyzing other products because is not dependent on $system$.
The cards on the bottom left of Figure~\ref{fig:flow_summary} represent a knowledge base of threats.

As described in Section~\ref{sec:asset:requirements} and depicted in \ARef*{alg:p2:threat_analysis:breakdown}, an attack's requirements can be described using the set of assets an adversary must gain capabilities over.
Papernot et. al.'s work in~\cite{papernotSoKSecurityPrivacy2018} presented a first step towards describing adversarial AI attacks based on their requirements but Section~\ref{sec:asset_caps} formalizes and standardizes asset types and capability ranges to the extent required for bottom-up analysis.
Section~\ref{sec:analysis:s3_attack:backdoor_example} describes the process for mapping the requirements for Backdoor attacks.
Table~\ref{tab:backdoor_breakdown} summarizes the identified requirements.


\subsubsection{
Example Breakdown: Backdoor attacks
}\label{sec:analysis:s3_attack:backdoor_example}
Consider, as an example, a backdoor attack~\cite{liBackdoorAttacksPretrained2021, sanchezvicarteDoubleCrossAttacksSubverting2021, turnerCleanLabelBackdoorAttacks} where the adversary’s goal is to circumvent a content filter; an image classifier for detecting inappropriate visual ads (e.g. those that promote racism or political extremism).
Backdoor attacks manipulate a classifier’s training process so that any input with an attacker-designed trigger pattern will be classified as benign.
A trigger pattern could be visual noise which is imperceptible to humans.
For a backdoor attack to be successful, an adversary must:
\begin{enumerate}
    \item Manipulate the Training Data, compromising its integrity.
    Some subset of the training data for the adversary’s target class – i.e. ads labeled as “benign” – must include the trigger pattern.

    \item Modify inference inputs so that they include the backdoor trigger, compromising their integrity.
    This is how the adversary exploits the backdoor.
    They overlay their chosen trigger pattern on non-benign ads, allowing those ads to circumvent the content filter and reach millions of internet users.
\end{enumerate}
Note that solely achieving either of these requirements is insufficient.
There will be no backdoor for the adversary to exploit if they cannot modify the training data.
Similarly, the attack will fail if the adversary successfully modifies the training data but cannot submit ads which include the trigger.
For example, because the adversary is not allowed, or cannot afford, to purchase the ad space.
Table~\ref{tab:backdoor_breakdown} and the top card on the bottom left of Figure~\ref{fig:flow_summary} depict the requirements to perform a backdoor attack.
The rest of this section describes how those requirements were determined for the training data asset type.

The Adversarial AI literature has identified many techniques which will allow an adversary to compromise the training data's integrity.
Four examples are listed below.
A description of the required capability has been included alongside each.
The first three describe techniques which allow an adversary to change benign training data.
That is, the adversary has obtained an Integrity capability and, implicitly, a Confidentiality capability.
The fourth technique is of particular note because the adversary has no access to benign training data:
They lack all Integrity or Confidentiality capabilities over benign training data.

\textbf{Example Techniques to Compromise Training Data:}
\begin{enumerate}
\item Apply trigger to benign examples in a public dataset.
Victims which consume this dataset will become vulnerable to the trigger pattern~\cite{shanNightshadePromptSpecificPoisoning2024, turnerCleanLabelBackdoorAttacks}.
It is often assumed that the adversary themselves hosts this dataset.
The adversary must also, somehow, ensure that the intended victim consumes the dataset.

\textbf{Capability: Make Limited Changes.} The adversary permanently modifies training data but modifications must be imperceptible to avoid detection.

\item Manipulate training data during collection or in storage.
An adversary with a foothold on a victim’s data collection infrastructure can apply the trigger to benign examples as they are collected, transformed, or encoded.
Similarly, an adversary could compromise the data at rest, in storage.
These modifications are permanent and persist across all future uses of the data – unless the victim is able to roll the dataset back to a previous version.

\textbf{Capability: Make Limited Changes.} The adversary permanently modifies training data but modifications must be imperceptible to avoid detection.

\item Manipulate training data during use.
The adversary transiently applies the trigger to benign data immediately before it is used for training.
For example, by compromising a data augmentation library~\cite{WhatDataAugmentation}.

As augmented data is typically not collected or reviewed, adversarial changes need not be stealthy.
Saving augmentation data effectively requires creating a unique copy of the dataset during every single epoch of training.
Doing so incurs massive storage costs and generates too much data for analysis to be feasible.

\textbf{Capability: Make Arbitrary Changes.} The adversary transiently modifies training data, which will not be reviewed, before it's consumed by the training algorithm.

\item Apply the trigger to benign-appearing data and inject it into a data collection or retraining pipeline.
For example, by leveraging techniques from~\cite{sanchezvicarteDoubleCrossAttacksSubverting2021, zhangOnlineDataPoisoning2020}.

This technique exemplifies an AI-specific way for adversaries to interact with data: by exploiting data collection or re-training pipelines.
This capability is also the minimal required capability, for a backdoor attack, because the adversary has no read or write access to any data besides their own.
Notably, they also also do not need a foothold on any victim infrastructure.

\textbf{Capability: Contribute.} The adversary adds to the training data but modifications must be imperceptible to avoid detection; e.g. during labeling.
\end{enumerate}

Table~\ref{tab:backdoor_breakdown} summarizes the relevant assets and \emph{minimum} capabilities an adversary must obtain over each.
Critically, obtaining \emph{stronger} capabilities enables an adversary to perform more powerful variants of backdoor attacks.
Using the minimum capabilities during their analysis allows a defender to identify the full range of backdoor attacks which are in-scope for their system.
Therefore, the minimum requirements are used when mapping threats to $AF$ in Figure~\ref{fig:flow_summary}.


\begin{table}[h]
\caption{This table summarizes Backdoor Attacks. Row 3 identifies the \emph{asset-centric attack requirements} for performing a backdoor attack.}
\begin{tabular}{p{0.18\columnwidth} | p{0.7\columnwidth}}
\toprule

\textbf{Backdoor Attacks} & Backdoor attacks allow an adversary to manipulate inference output.

\\ \midrule

Technique    & Attacker manipulates classifier's training process so that any input with an attacker-designed trigger pattern will classify to an attacker-chosen target class.

\\ \midrule

Attack Requirements & Adversary must compromise the \textbf{Training Dataset} and \textbf{Inference Inputs}.
\begin{enumerate}
    \item \textbf{Training Dataset: Contribute.} Adding malicious data to the training set.
    \item \textbf{Inference Inputs: Make Limited Changes.} Adding the backdoor trigger to inputs, leaving them otherwise unchanged.
\end{enumerate}
\\ \midrule

Impact       & 
Adversary can \emph{Influence} \textbf{Output Classification} but only for inputs which include the backdoor trigger and only to convert it into the target classification.
\\
\bottomrule
\end{tabular}
\label{tab:backdoor_breakdown}
\end{table}

\subsection{
Mapping an attack into a product (Phase 2)
}\label{sec:analysis:s4_mapping}
\begin{figure*}[!t]
    \centering
    \includegraphics[width=\textwidth]{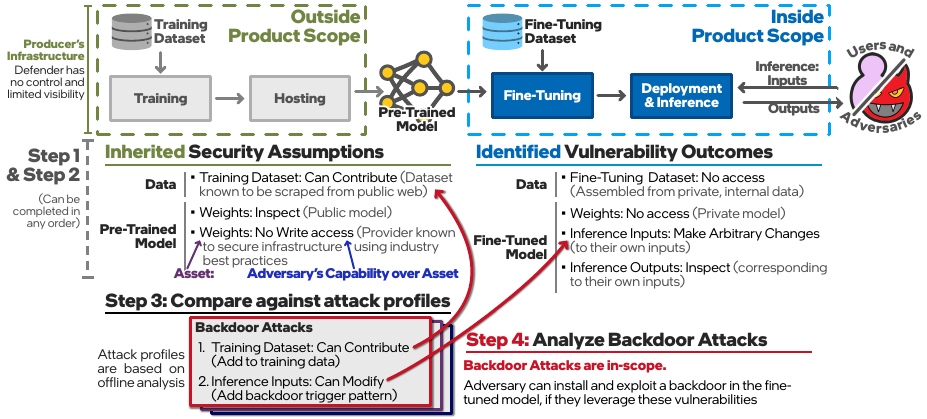}
    \caption{
    This figure depicts an asset-centric analysis across an example AI lifecycle.
    The top left lists various security assumptions, including justifications, made about the pre-trained model being consumed.
    The steps necessary to train are outside product scope, preventing defenders from directly inspecting them.
    The top right lists various vulnerabilities identified during security analysis of the steps which are in product scope.
    The bottom depicts how a defender can combine the insights from an attack profile with the security assumptions and identified vulnerabilities to determine if, and how, a specific attack is possible.
    }
    \label{fig:flow_summary}
\end{figure*}

The second step in phase 2 (\ARef*{alg:overall:mapThreats})
compares the aggregated capabilities in $AF$ against the threats in $KB$ to identify in-scope threats.
Having broken down an AI attack as described in Section~\ref{sec:analysis:s3_attack} and added it into the knowledge base $KB$, defenders determine whether that attack could be enabled by obtaining any combination(s) of the capabilities identified in $AF$.
As depicted in \ARef*{alg:p2:check_requirements}, this is done by comparing the capabilities an adversary could gain within the AI's lifecycle -- including implied capabilities (Section~\ref{sec:asset_caps:caps:implied}) -- against the attack requirements.
If the attack's requirements can be met by some combination of vulnerabilities, the attack is deemed to be in scope and added into $TM$.
This analysis is caricatured by the red arrows in Figure~\ref{fig:flow_summary},
The defender may identify \emph{multiple} vulnerabilities, across the stack and the distributed pipeline, which provide the required capability.
Knowing that the adversary could exploit \emph{any} of these vulnerabilities or assumptions, as a link in their chain of attacks, facilitates the defender’s risk analysis.

\begin{algorithm}
\caption{Phase 2: Mapping Threats Based on Adversary Footprint (Scalable Analysis)}
\label{alg:p2}
\begin{algorithmic}[1]
\Procedure{MapThreats}{$AF, KB$} \label{alg:p2:map_threats}
    \State $inScope \gets \emptyset$
    \State $outScope \gets \emptyset$ 
    
    \For{each threat $t$ in $KB$} 
        \If{RequirementsMet($t.requirements, AF$)}
            \State $vectors \gets$ IdentifyAttackVectors($t, AF$)
            \State Add $(t, vectors)$ to $inScope$ 
        \Else
            \State $just \gets$ ExplainUnmetRequirements($t, AF$)
            \State Add $(t, just)$ to $outScope$ 
        \EndIf
    \EndFor
    
    \State \Return $(inScope, outScope)$ 
\EndProcedure

\Procedure{AnalyzeNewThreat}{$newThreat, AF$} \label{alg:p2:threat_analysis}
    \State Breakdown $newThreat$ into its requirements\label{alg:p2:threat_analysis:breakdown}
    \State Add $newThreat$ to knowledge base
    \If{RequirementsMet($newThreat.requirements, AF$)}
        \State \Return "In scope" with attack vectors
    \Else
        \State \Return "Out of scope" with justification
    \EndIf
\EndProcedure

\Procedure{RequirementsMet}{$requirements, AF$} \label{alg:p2:check_requirements}
    \For{each requirement $r$ in $requirements$}
        \State $met \gets$ false
        \For{each capability $c$ in $AF$}
            \If{$c$ satisfies $r$}
                \State $met \gets$ true
                \State \textbf{break} 
            \EndIf
        \EndFor
        \If{not $met$}
            \State \Return false
        \EndIf
    \EndFor
    \State \Return true
\EndProcedure
\end{algorithmic}
\end{algorithm}

Notably, because $TM$ is generated using $AF$, the defender does not need in depth knowledge of each vulnerability or its originating system architecture when implementing \ARef{alg:p2}.
The defender may even choose not mitigate a threat at the “root cause” vulnerabilities in $AF$.
For example, a defender may opt to mitigate the risks of a backdoor attack (Table~\ref{tab:backdoor_breakdown}) by implementing a consensus system instead of performing high-cost and high-complexity model retraining.
This mitigation instead opts to increases confidence in AI outputs by augmenting them with information from a non-AI source – such as a LIDAR sensor in an autonomous vehicle – or a second model.

\section{Enterprise RAG: A Secure Retrieval Augmented Generation Architecture}\label{sec:enterprise_rag_bg}

\begin{figure}
    \centering
    \includegraphics[width=1\linewidth]{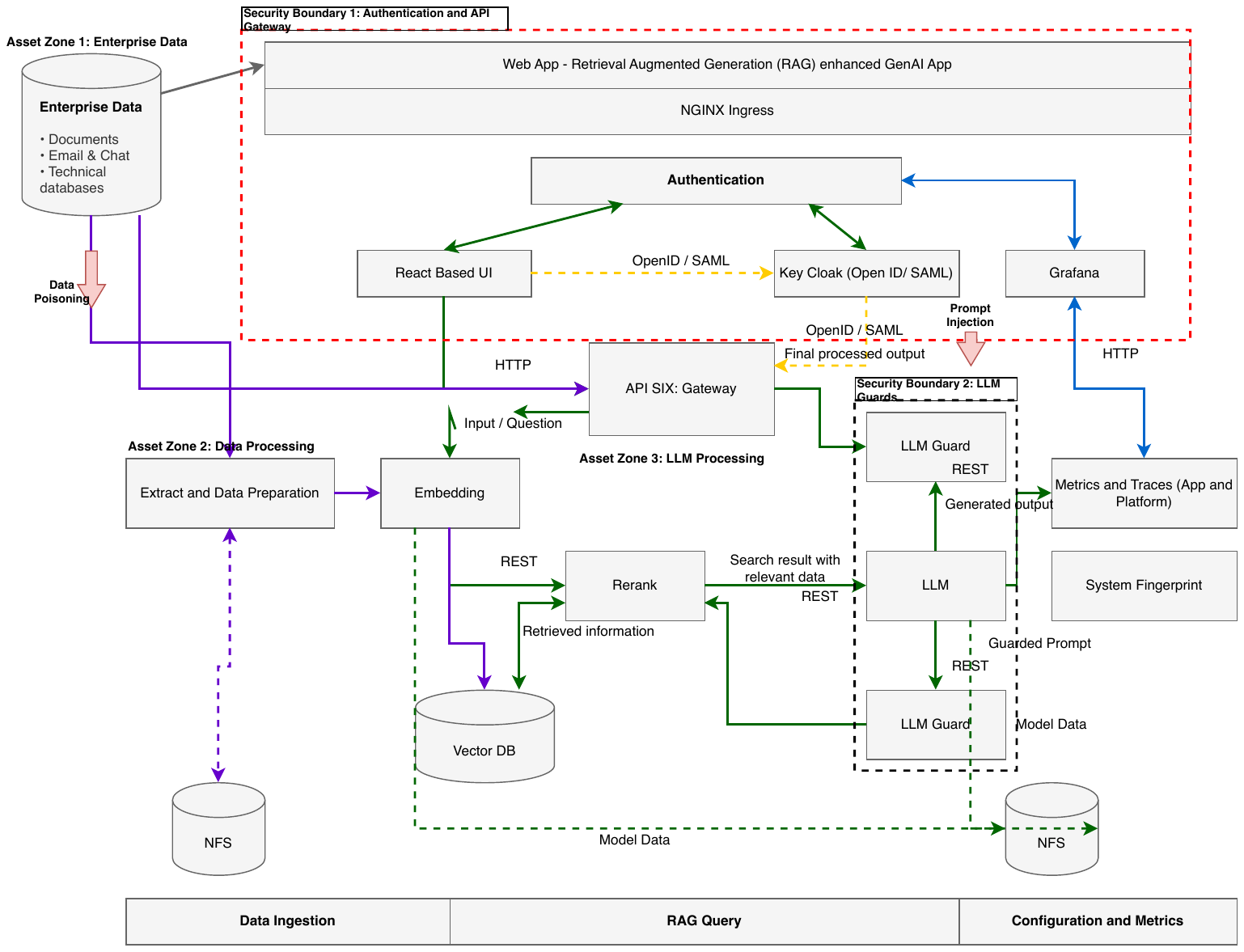}
        \caption{Enterprise RAG Architecture with Asset-Centric Security Analysis. The diagram illustrates a complete Retrieval Augmented Generation system with four distinct asset zones: (1) Enterprise Data (top-left), containing source documents and databases; (2) Data Processing (bottom-left), including extraction, embedding, and vector storage components; (3) LLM Processing (center-right), featuring reranking, LLM, and guard components; and (4) Monitoring (far-right) with metrics and system fingerprinting. Security boundaries (authentication layer and LLM guards) protect critical assets. Red arrows indicate potential attack vectors: data poisoning, prompt injection, vector manipulation, and retrieval tampering. This asset-centric approach enables systematic vulnerability identification across distributed components, regardless of deployment context. Each asset is evaluated for adversarial capabilities concerning confidentiality, integrity, and availability.}

    \label{fig:rag-depl}
\end{figure}

The Enterprise RAG architecture implements a secure Retrieval Augmented Generation system for enterprise deployments. As illustrated in Figure~\ref{fig:rag-depl}, the system follows an asset-centric security approach, where each component is treated as a potential attack surface with specific security controls~\cite{entrag2024}.

The architecture comprises three primary functional domains: Data Ingestion, RAG Query Processing, and Configuration/Metrics. At the entry point, an NGINX Ingress layer provides initial request handling, with all traffic flowing through a comprehensive authentication layer that leverages OpenID/SAML protocols. A React-based UI interfaces with users, while an API SIX Gateway mediates all communication between frontend components and backend services. This gateway serves as a critical security boundary, enforcing access controls and traffic validation.

The data flow begins with enterprise data sources (documents, email, chat, and technical databases) being processed through an extraction and data preparation pipeline. The processed data is then embedded and stored in a secure Vector Database. When users submit queries, the system enhances standard LLM processing with a retrieval mechanism that incorporates relevant contextual information. This involves several secured microservices: Embedding, Rerank, and LLM components, with LLM Guard modules providing content filtering and safety validation at two critical points - before prompt construction and after generation.

\section{Applying Asset-Centric Threat Modeling to Enterprise RAG}\label{sec:case_study}

To demonstrate the application of asset-centric threat modeling to AI systems, we onboard the Enterprise RAG~\cite{entrag2024} (Section~\ref{sec:enterprise_rag_bg}) architecture into this framework. Following the asset-centric methodology we begin by identifying the key AI assets within the system and the potential adversarial capabilities over these assets.

\subsection{Asset Identification and Capability Mapping}

For our Enterprise RAG system, we identify the following critical AI assets:

\begin{itemize}
    \item \textbf{Enterprise Data}: Source documents containing sensitive corporate information
    \item \textbf{Embedding Model}: Transforms raw text into vector representations
    \item \textbf{Vector Database}: Stores embeddings with semantic search capabilities
    \item \textbf{LLM}: Large language model for generating responses
    \item \textbf{Fine-tuning Dataset}: Data used to adapt the LLM to enterprise context
    \item \textbf{User Queries}: End-user inputs that could be manipulated
    \item \textbf{Generated Responses}: Output text that must maintain accuracy and safety
\end{itemize}

For each asset, we map the potential adversarial capabilities. For instance, with the Vector Database:
\begin{itemize}
    \item \textbf{Read Access}: An adversary could extract sensitive embeddings
    \item \textbf{Write Access}: An adversary could poison the database with malicious vectors
    \item \textbf{Contribute Data}: Limited to adding new entries without modifying existing ones
\end{itemize}

\subsection{Vulnerability Analysis and Threat Contextualization}

Through security analysis, we identified several potential vulnerabilities in our implementation:
\begin{itemize}
    \item \textbf{API Gateway}: Potential for request manipulation if authentication is bypassed
    \item \textbf{LLM Guard}: Risk of prompt injection if guardrails are insufficient
    \item \textbf{Vector DB}: Potential for embedding extraction if access controls are compromised
\end{itemize}

By correlating these vulnerabilities with our asset-capability mapping, we can contextualize specific attacks. For example, a jailbreaking attack on our Enterprise RAG would require:

\begin{enumerate}
    \item Capability over \textbf{User Queries}: Ability to craft specially formatted inputs
    \item Capability over \textbf{LLM Guard}: Ability to bypass prompt filtering mechanisms
    \item Capability over \textbf{Generated Responses}: Ability to receive unfiltered LLM output
\end{enumerate}

Our security analysis identified that while the system has strong protections for the Vector Database and enterprise data sources, the LLM Guard component could potentially be bypassed through sophisticated prompt engineering. This leads to a focused mitigation strategy targeting this specific vulnerability rather than implementing broad, untargeted security measures. 

\begin{table}[htbp]
    \centering
    \caption{Asset-Centric Security Analysis of Enterprise RAG Architecture}
    \begin{tabular}{p{1cm}p{2cm}p{2cm}p{2cm}}
    \toprule
    \textbf{Asset Zone} & \textbf{Key Assets} & \textbf{Adversarial Capabilities} & \textbf{Security Controls} \\
    \midrule
    Enterprise Data & Source documents, Email, Technical databases & Read (data theft), Write (data poisoning) & Access controls, Data validation, Integrity checks \\
    \midrule
    Data Processing & Embedding model, Vector DB, NFS & Read (model stealing), Write (embedding poisoning), Contribute (fake data) & Embedding validation, Anomaly detection, Distribution checks \\
    \midrule
    LLM Processing & Reranker, LLM, LLM Guards & Read (prompt leakage), Execute (jailbreaking) & Input sanitization, Output filtering, Context boundaries \\
    \midrule
    Monitoring & Metrics, System Fingerprint & Write (log tampering), Read (security insights) & Immutable logs, Real-time monitoring \\
    \bottomrule
    \end{tabular}
    \label{tab:asset-security}
\end{table}

\section{Agent-Centric Example: Automated Rust Assistant}

To further illustrate our asset-centric methodology in a practical context, we consider Automated Rust Assistant, an AI-powered development agent designed to autonomously write, debug, and optimize Rust code. This example demonstrates how our approach can be applied to agentic AI systems that interact with development ecosystems.

\subsection{System Overview}

Automatic Rust Assistant is an autonomous AI agent that:
\begin{enumerate}
    \item Receives high-level programming tasks from users
    \item Generates Rust code to fulfill these requirements
    \item Executes \texttt{cargo build} commands to compile the code
    \item Analyzes compiler errors and warnings
    \item Debugs issues using cargo-debugger and VSCode integration
    \item Optimizes code based on performance metrics
    \item Commits changes to version control systems
\end{enumerate}

The agent has significant capabilities beyond typical AI assistants, including:
\begin{itemize}
    \item Direct access to file systems for reading and writing Rust project files
    \item Ability to execute shell commands (restricted to cargo ecosystem)
    \item Integration with the Debug Adapter Protocol (DAP) used by VSCode to programmatically set breakpoints and receive debugging information
    \item Access to version control systems to commit optimized code
\end{itemize}

\subsection{Asset Identification}

Following our asset-centric methodology, we identify the key assets in Automated Rust Assistant:

\subsubsection{Code Generation Model}
\begin{itemize}
    \item Pre-trained on Rust codebases and documentation
    \item Fine-tuned on high-quality, production Rust code
\end{itemize}

\subsubsection{Error Analysis Model}
\begin{itemize}
    \item Trained to interpret and resolve Cargo build errors
    \item Fine-tuned on real-world debugging sessions
\end{itemize}

\subsubsection{Execution Environment}
\begin{itemize}
    \item Shell access (limited to cargo commands)
    \item File system access (restricted to project directories)
    \item Debug Adapter Protocol (DAP) client capabilities for interacting with debuggers
\end{itemize}

\subsubsection{Runtime Behavior Dataset}
\begin{itemize}
    \item Collection of runtime behavior and performance metrics
    \item Used to guide optimization decisions
\end{itemize}

\subsubsection{User Project Codebase}
\begin{itemize}
    \item Source code files the agent can modify
    \item Dependencies and build configurations
\end{itemize}

\subsubsection{Cargo Credentials}
\begin{itemize}
    \item Potential access to private registry credentials
    \item Could enable publishing or installing packages
\end{itemize}

\subsection{Vulnerability Analysis}

Using our asset-centric approach, we identify several potential vulnerabilities:
\begin{table}[h]
\caption{Asset-Centric Vulnerability Analysis for AutoRust: This table maps key assets to potential adversarial capabilities and corresponding security controls.}
\begin{tabular}{p{0.18\columnwidth} | p{0.28\columnwidth} | p{0.42\columnwidth}}
\toprule
\textbf{Asset} & \textbf{Adversarial Capability} & \textbf{Security Controls} \\
\midrule
Code Generation Model & 
Manipulate model to produce malicious or vulnerable code through specially crafted prompts & 
\begin{itemize}
    \item Input validation and sanitization
    \item Security-focused model fine-tuning
    \item Output scanning for known vulnerability patterns
\end{itemize} \\
\midrule
Debug Adapter Protocol & 
Exploit DAP to access memory contents beyond intended scope & 
\begin{itemize}
    \item Restricted DAP client with limited command scope
    \item Runtime monitoring for suspicious memory requests
    \item Sandboxed debugging environment
\end{itemize} \\
\midrule
Cargo Dependency Resolution & 
Inject malicious packages through dependency confusion & 
\begin{itemize}
    \item Enforce private registry specifications
    \item Checksum verification for all dependencies
    \item Dependency lockfiles with integrity checks
\end{itemize} \\
\midrule
Shell Command Interface & 
Execute arbitrary system commands through command injection & 
\begin{itemize}
    \item Restricted command whitelist
    \item Execution in isolated container
    \item Command argument sanitization
\end{itemize} \\
\midrule
File System Access & 
Read/write sensitive files beyond project scope & 
\begin{itemize}
    \item Path traversal prevention
    \item Read-only mountpoints where possible
    \item File access auditing
\end{itemize} \\
\bottomrule
\end{tabular}
\label{tab:asset_vulnerability_mapping}
\end{table}

\subsection{Threat Modeling Example: Dependency Confusion Attack}

Applying our asset-centric methodology, we analyze how a dependency confusion attack could be implemented against AutoRust:

\subsubsection{Attack Requirements:}
\begin{enumerate}
    \item \textbf{User Project Codebase}: Write access to Cargo.toml (to modify dependencies)
    \item \textbf{Execution Environment}: Ability to execute \texttt{cargo build} (to resolve dependencies)
    \item \textbf{Code Generation Model}: Ability to influence code generation to include vulnerable patterns
    \item \textbf{Cargo Credentials}: Read access to determine private registry configurations
\end{enumerate}

\subsubsection{Attack Implementation:}
\begin{enumerate}
    \item Attacker crafts a prompt that instructs the agent to use an internal dependency
    \item Agent updates Cargo.toml with the dependency name but without specifying the private registry
    \item When \texttt{cargo build} executes, Cargo first checks the public registry (crates.io)
    \item If the attacker has published a malicious package with the same name on the public registry, it gets downloaded instead of the internal package
    \item The malicious dependency executes in the context of the project
\end{enumerate}

\subsubsection{Asset-Centric Analysis:}
\begin{itemize}
    \item The vulnerability stems from the agent's capabilities over multiple assets: write access to project files and ability to execute cargo commands
    \item The attack chains these capabilities together to compromise the build process
    \item Traditional security would focus on the execution environment, while our approach reveals the full attack path across multiple assets
\end{itemize}

\begin{table}[h]
\caption{This table summarizes the Dependency Confusion Attack against AutoRust. Row 3 identifies the \emph{asset-centric attack requirements} needed to execute this attack.}
\begin{tabular}{p{0.18\columnwidth} | p{0.7\columnwidth}}
\toprule
\textbf{Dependency Confusion Attack} & An attack that exploits package resolution mechanisms to inject malicious code into the development environment.
\\ \midrule
Technique    & Attacker publishes a malicious package with the same name as an internal dependency to a public registry. When dependency resolution occurs, the malicious package is downloaded instead of the legitimate internal package.
\\ \midrule
Attack Requirements & Adversary must compromise \textbf{User Project Codebase}, \textbf{Execution Environment}, and have knowledge of \textbf{Cargo Credentials}.
\begin{enumerate}
    \item \textbf{User Project Codebase:} Write access to Cargo.toml to modify dependencies.
    \item \textbf{Execution Environment:} Ability to trigger \texttt{cargo build} execution for dependency resolution.
    \item \textbf{Cargo Credentials:} Knowledge of internal dependency names without registry specifications.
\end{enumerate}
\\ \midrule
Impact & 
Malicious code executes within the project's context with the same privileges as legitimate dependencies, potentially stealing sensitive data or corrupting the development process.
\\
\bottomrule
\end{tabular}
\label{tab:dependency_confusion_breakdown}
\end{table}

\subsection{Mitigations}

Based on our methodology, we identify asset-centric mitigations:

\subsubsection{Execution Environment Constraints}
\begin{itemize}
    \item Implement a custom cargo wrapper that always enforces private registry use
    \item Add post-execution validation to detect unexpected registry sources
    \item Create a secure build sandbox isolated from production environments
\end{itemize}

\subsubsection{Code Generation Model Guardrails}
\begin{itemize}
    \item Add specific validation for dependency-related operations
    \item Train the model to always specify complete registry information
    \item Implement pattern recognition for potentially unsafe dependency changes
\end{itemize}

\subsubsection{Integration Boundaries}
\begin{itemize}
    \item Require human approval for changes to dependency configurations
    \item Create checksum verification for critical project files
    \item Implement multi-stage verification before executing cargo commands
\end{itemize}

\subsubsection{Debug Adapter Protocol Security}
\begin{itemize}
    \item Implement a restricted Debug Adapter Protocol client with limited command scope
    \item Add runtime monitoring for suspicious memory inspection requests
    \item Create validation policies for debugger interactions
    \item Develop a sandboxed debugging environment with memory access controls
\end{itemize}

This example demonstrates how our asset-centric approach provides a comprehensive framework for analyzing complex security challenges in AI agents. By focusing on the assets and the adversarial capabilities over those assets, defenders can identify vulnerabilities that might be missed through traditional security analysis.

\begin{figure}
    \centering
    \includegraphics[width=1\linewidth]{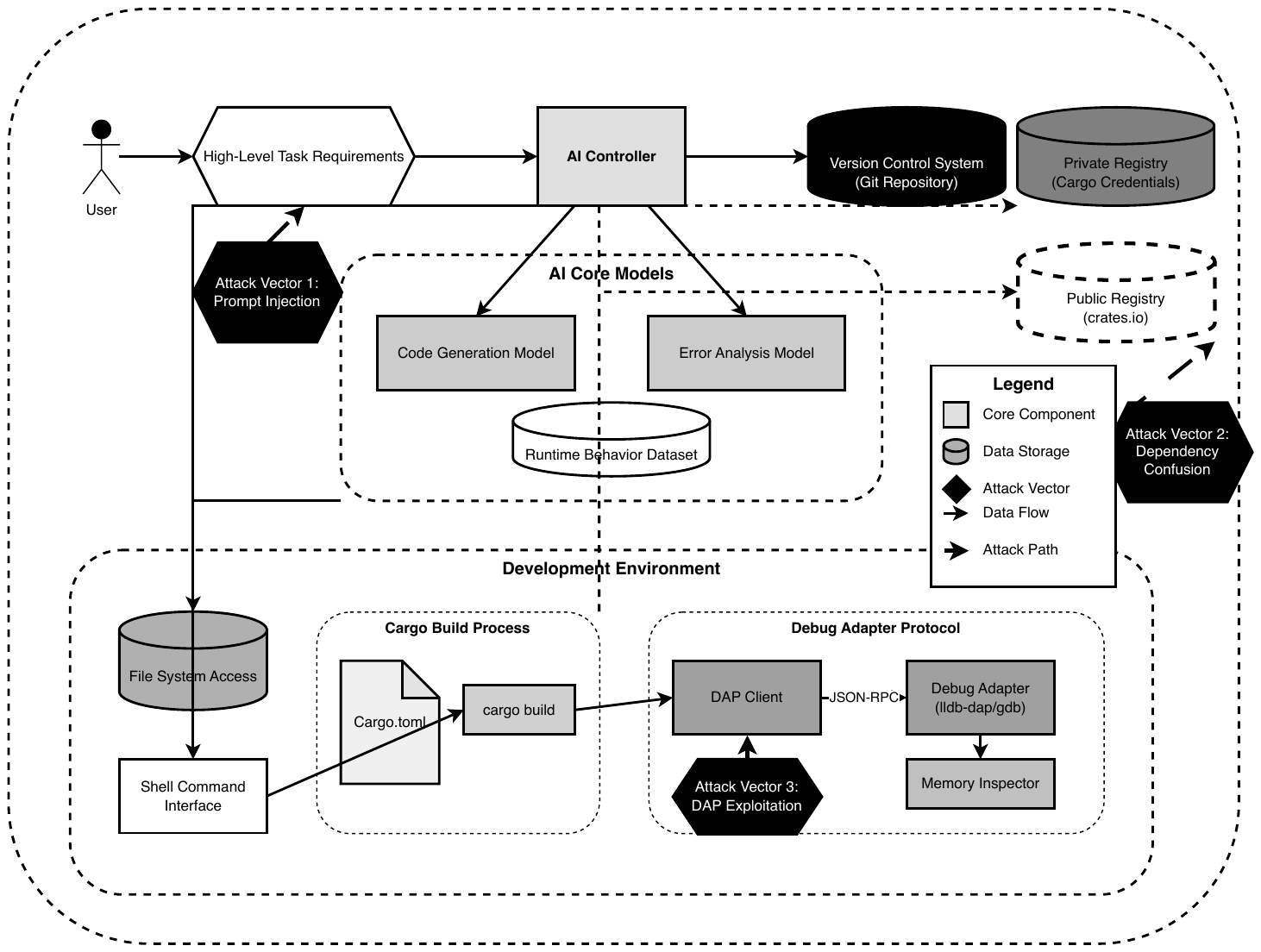}
        \caption{Asset-centric view of the AutoRust system architecture showing key components, data flows, and attack vectors. The system consists of an AI Controller that orchestrates multiple subsystems: (1) AI Core Models (including Code Generation and Error Analysis models), (2) Development Environment (with file system access, shell command interface, Cargo build process, and Debug Adapter Protocol integration), and (3) external registries and repositories. Three primary attack vectors are highlighted: prompt injection targeting user input, dependency confusion exploiting registry resolution, and Debug Adapter Protocol exploitation targeting debugging interfaces. Solid arrows represent normal data flows, while dashed red arrows indicate potential attack paths. This architecture illustrates how autonomous AI agents that interact with development tools present unique security challenges that span multiple asset boundaries, requiring comprehensive asset-centric threat modeling.}
    \label{fig:enter-label}
\end{figure}

\subsection{Extension: Debug Adapter Protocol Attack}

To further illustrate the power of asset-centric analysis, consider a more sophisticated attack that leverages AutoRust's integration with the Debug Adapter Protocol (DAP) used by debuggers like LLDB and GDB:

\subsubsection{Attack Scenario}
An attacker with the goal of accessing sensitive information from the debugging session could:

\begin{enumerate}
    \item Craft a prompt requesting AutoRust to implement a Rust program with legitimate debugging needs
    \item Include specific code patterns designed to trigger memory exposure bugs when debugged
    \item Manipulate the Debug Adapter Protocol messages exchanged between the DAP client (AutoRust) and the debug adapter
    \item Cause the debugger to reveal memory contents beyond the intended scope
    \item Extract sensitive data (e.g., encryption keys, authentication tokens) through debug variables
\end{enumerate}

\subsubsection{Technical Context}
The Debug Adapter Protocol is a JSON-based protocol used by many IDEs (including VSCode) to communicate with language-specific debuggers. AutoRust interacts with this protocol programmatically by:

\begin{itemize}
    \item Launching a debug adapter process (e.g., \texttt{lldb-dap} for Rust)
    \item Sending JSON-RPC requests to control the debugging session
    \item Receiving events and responses containing debugging information
    \item Setting breakpoints, examining variables, and evaluating expressions
\end{itemize}

\subsubsection{Asset-Centric Analysis}
\begin{itemize}
    \item \textbf{Debug Adapter Protocol}: The primary communication channel being manipulated
    \item \textbf{Debug Adapter Process}: The target debug engine (LLDB/GDB) processing potentially malicious commands
    \item \textbf{Code Generation Model}: Manipulated to create code with specific debugging vulnerabilities
    \item \textbf{Memory Inspector Capabilities}: Used to access memory regions outside the intended scope
    \item \textbf{User Project Environment}: Contains sensitive data that could be exposed
\end{itemize}

This more complex attack demonstrates how asset-centric analysis helps defenders reason about cross-boundary attacks that span the AI agent, the debugging protocol, and the underlying memory management system—a vulnerability chain that traditional security analysis might miss, particularly when dealing with AI agents that programmatically control development tools.

\section{Related Work}\label{sec:related}
An asset-centric approach is highly complementary to existing AI threat modeling frameworks because existing frameworks take a top-down perspective while an asset-centric approach is bottom-up.
In practice, defenders will benefit from blending the two perspectives.


\subsubsection{Perspective for Analysis}
Today's AI threat modeling frameworks map the space of potential AI attacks, broadly describing the techniques adversaries might use to perform each one~\cite{papernotSoKSecurityPrivacy2018, OWASPTopTen, ATLASMatrixMITRE}.
MITRE’s ATLAS framework~\cite{ATLASMatrixMITRE}, for example, categorizes attack techniques based on their role within MITRE's attack lifecycle; allowing defenders to reason about a specific attack based on the steps an adversary takes to achieve that attack.
Defenders use this approach to contextualize a specific attack within their product by individually mapping every step the adversary takes, from Reconnaissance and Resource Development through to Exfiltration and Impact. 
This is a top-down perspective because the defender’s analysis maps the attack onto their product.

A defender seeking to prevent an adversary from manipulating the output of their AI, for example, would need to consider every possible attack technique which could lead to output manipulation.
As potential techniques span across the AI's lifecycle, defenders must individually contextualize many attack types within their product.
This approach, therefore, does not scale with a quickly growing and evolving attack space.
It also does not inherently leverage the insights from a defender's existing security analysis.

Defenders cannot quickly determine whether a vulnerability identified in their threat model could enable an AI attack.
That is, whether it could serve as a step in an attack’s lifecycle.
Contextualizing a step in an attack's lifecycle could require reanalyzing every vulnerability identified in the threat model, to determine potential relevance.
This is a particularly challenging task when reason about the ways in which conventional hardware or security vulnerabilities can facilitate AI threats.

Asset-centric analysis helps invert this perspective.
It’s \emph{centered} around allowing defenders to determine whether the vulnerabilities they’ve already identified in their threat model can serve as steps in an AI attack’s lifecycle.
That is, by allowing them to identify how a vulnerability impacts AI assets and reason about whether that impact enables subsequent attacks.
Importantly, defenders can reason about many attacks by \emph{reusing} their analysis.
Finally, it is more accessible for product domain and security experts – who may not be AI security experts, but do have asset-centric threat modeling experience.
Their security analysis can scale with a quickly evolving, and often nuanced, AI threat landscape by reasoning about threats at an asset level.

\subsubsection{Analysis Across System Boundaries}
Defenders lack a straightforward way to apply today's frameworks to AI systems distributed across multiple actors.
Consuming AI assets from external, producing, actors – such as datasets or pretrained models – allows developers to
leverage state of the art AI assets without the costs of creating them from scratch.
However, due to their proprietary nature, defenders often lack visibility into the systems which produced those assets.
Defenders are unable to contextualize attacks within those systems, as required by a top-down approach.
This prevents defenders from leveraging today's frameworks to identify and quantify the security assumptions they make about assets consumed from external, closed box, systems.

Defenders can overcome this challenge through a \emph{boundary based} approach, enabled by asset-centric analysis.
This approach does not require defenders to reason about individual vulnerabilities within systems they lack visibility over.
That is, defenders can identify, quantify, and justify their security assumptions about consumed assets at system boundaries – at the point where assets transition across contexts.
As described in Section~\ref{sec:analysis:s2_assumptions}, there are many ways for defenders to justify constraints on their security assumptions.
They might, for example, leverage provenance information, the producer's reputation, or properties about the asset itself.
Conversely, producers could leverage an asset-centric perspective to communicate the security guarantees they provide – without needing to open their systems.
Even AI-enabling products – such as processors, accelerators and GPUs, or cloud infrastructures – could leverage this perspective to describe the security guarantees they provide for AI assets.

\subsubsection{Complementary Analysis}
Top-down and bottom-up perspectives are not mutually exclusive.
Ultimately, defenders will benefit most from incorporating both top-down and bottom-up perspectives into their security analysis.
Existing frameworks, including ATLAS and~\cite{mcgrawARCHITECTURALRISKANALYSIS,khawajaDatabricksAISecurity2024}, could also augment their attack knowledge bases with asset-centric information.
Defenders could then leverage these knowledge bases for whichever perspective – or combination of perspectives – best fit their unique needs.

\section{Ongoing Work and Next Steps}\label{sec:future}
We have identified, and are pursuing, various opportunities to facilitate the adoption and automation of asset-centric analysis for threat modeling AI.
The security analysis in Phase 1 can be facilitated by standardizing the types of AI assets, the range of potential capabilities adversaries can obtain over them (Sections~\ref{sec:analysis:s1_vulns} and~\ref{sec:analysis:s2_assumptions}).
The threat analysis in Phase 2 can be facilitated by building and augmenting AI attack knowledge bases with asset-centric information.
Defenders can reference such knowledge bases instead of reanalyzing attacks (Section~\ref{sec:analysis:s3_attack}).
Finally, threat mapping can be automated using these knowledge bases (Section~\ref{sec:analysis:s4_mapping}).
These opportunities are, in large part, enabled by standardizing the types of AI assets and mapping out the range of possible capabilities adversaries may obtain over each.

While our formalized knowledge base and standardization efforts continue to develop, organizations may consider taking initial steps toward this methodology. Security teams could begin by identifying key AI assets within their existing environments and considering how these assets might be affected by current threats. Even a simplified version of the asset-capability mapping can provide valuable insights when integrated into existing security processes. By bringing together security practitioners and AI specialists for collaborative threat analysis sessions, organizations can build the cross-functional understanding needed for effective AI security, laying groundwork that will align well with more comprehensive asset-centric approaches as they mature.
 
\subsection{Standardizing Assets}
As a foundation for facilitating, optimizing, and automating asset-centric analysis, we have enumerated the different types of AI assets and have mapped out the extents to which each can be compromised.
Just like for non-AI assets, the range of potential capabilities that an adversary can gain over each type of AI asset can be statically mapped out for each asset type.
Today, defenders often denote whether a vulnerability would compromise an asset's Confidentiality, Integrity, or Availability.
These ranges are particularly interesting for AI assets
because adversaries can gain previously unseen, nuanced, capabilities over them.
One such capability, discussed in the example within Section~\ref{sec:analysis:s3_attack}, is an adversary’s ability to \emph{add} data to an AI’s training or re-training datasets but lack the ability to read or write the rest of the dataset.
Standardization also makes threat models easier to write, interpret, and reuse.
Especially for those who are product domain or security experts but not AI security experts.
It also facilitates reuse of security assumptions and threat analysis.

\subsection{Reusing Security Assumptions}
Security assumptions about a consumed asset can be reused across different products, because those assumptions are not specific to the defender's product.
Defenders' security assumptions about that asset are based on their assumptions about the \emph{producer's} systems (Sectio~\ref{sec:analysis:s2_assumptions}).

Consider, for example, two products which consume the same pre-trained model.
Security assumptions about the pre-trained model are based on how that model is \emph{created} – they are not based on how that model is \emph{used} once consumed.
It's the \emph{interpretation} of those assumptions – i.e. their potential impact – which is product-specific.
As with conventional hardware and software security, the security requirements of two products can be significantly different.
Even if they share common components.
Critically, security assumptions must be based on up-to-date justifications for them to be reused.
However, it’s easier for defenders to verify that previous justifications still hold than it is for them to repeat the analysis from scratch.

\subsection{Automating Threat Analysis}
We are building a knowledge base of AI attacks augmented with asset-centric information.
To maximize re-usability, this knowledge base uses our standardized asset types.

Defenders could \emph{programmatically} compare each entry in this database – rather than manually – against the potential capabilities they identified in $AF$ (Section~\ref{sec:analysis:s4_mapping}).

\section{Conclusion}\label{sec:conclusion}
Security analysis for AI must overcome three key challenges: (1) increasingly distributed development and deployment pipelines across disjoint infrastructures (2) the need for holistic evaluation of vulnerabilities across system boundaries, and (3) limited visibility into external components that AI systems increasingly depend upon.
Defenders face a complex task when identifying vulnerabilities in these distributed environments. They must not only locate individual vulnerabilities within separate contexts but also understand how adversaries might chain these vulnerabilities across system boundaries. 
This challenge is compounded when systems incorporate external AI assets like pre-trained models, where defenders lack visibility into the development processes that created these components.

Current threat modeling frameworks employ a top-down approach that focuses on contextualizing specific attacks within product boundaries. This approach becomes increasingly untenable as AI systems grow more complex for two reasons: it requires comprehensive visibility across all development contexts, and it forces defenders to repeatedly reanalyze their entire system for each new attack pattern. As a result, these approaches cannot efficiently scale to address the rapidly evolving AI threat landscape.



This work presents a bottom-up, asset-centric, analysis that can be used to overcome these challenges.
This analysis is implemented in two phases.
First, defenders identify and quantify all potential adversarial influence over AI assets; both for assets within their product scope and assets consumed from external, often closed-box, systems.
Second, defenders analyze AI attacks – from an asset-centric perspective – so that they can contextualize those attacks within their product.
This process allows defenders to determine whether the attacks are feasible within their system and to identify the root-cause vulnerabilities that enabled those attacks.
Importantly, this methodology scales with a growing attack space because defenders \emph{reuse} their analysis from the first phase when they're contextualizing attacks in the second phase.
That is, unlike with top-down approaches, defenders need not reanalyze their product when evaluating each attack.

This work also describes ongoing and future work which will facilitate integrating and automating this methodology.
Specifically, and externally to this work, we have identified the different possible types of AI assets and mapped out the different degrees of control and/or visibility an adversary can gain over each one.
Combined with the methodology presented in this work, standardizing AI assets will
allow the security assumptions about consumed assets to become portable across products,
allow AI security experts to develop and augment AI attack knowledge bases with asset-centric information,
and allow defenders to automate the mapping of attacks from those knowledge bases onto defenders' product-specific context.

Ultimately, defenders will benefit most from \emph{complementing} top-down analysis – when it's feasible – with bottom-up analysis.
Top-down and bottom-up perspectives are not mutually exclusive and each has unique benefits.
This work takes the first step towards enabling both perspectives by presenting defenders with the first bottom-up approach for threat modeling AI.

\bibliographystyle{plain}
\bibliography{99_ai_tm}

\begin{table*}
\centering
\caption{
Examples of components, dependencies with other asset types, and relationships with other models or agents.
}
\label{tab:asset-types}
\begin{tblr}{
  width = \linewidth,
  colspec = {Q[135]Q[268]Q[268]Q[268]}, 
  hlines,
  vlines,
  hline{1,10} = {-}{0.08em},
}
\textbf{Asset Type} & \textbf{Components} & \textbf{Dependencies} (other Asset Types) & \textbf{Relationships} (other Models or Agents) \\

\textbf{Inputs.} & {\textit{User Prompt}:\\
\labelitemi\hspace{\dimexpr\labelsep}User Input.\\
\labelitemi\hspace{\dimexpr\labelsep}Prompt Context.\\
\labelitemi\hspace{\dimexpr\labelsep}Prompt Augmentations.\\
\labelitemi\hspace{\dimexpr\labelsep}Tool or API descriptions.\\
\labelitemi\hspace{\dimexpr\labelsep}Tool or API results.} &
Not inherently dependent on other asset types. &
Oftentimes, the Inputs to a model or agent are the Output of another AI. Example transitions are listed below, under the Output Asset Type. \\

\textbf{Outputs.} & {\textit{LLM Generated Content}:\\
\labelitemi\hspace{\dimexpr\labelsep
}Error messages.\\
\labelitemi\hspace{\dimexpr\labelsep}Tool or API actions or requests.\\
}
& {\labelitemi\hspace{\dimexpr\labelsep}Inputs\\
\labelitemi\hspace{\dimexpr\labelsep}Weights, Activations} & {Common~\textbf{\textbf{Output}}~$\rightarrow$ ~\textbf{\textbf{Input}}~transitions:\\
\labelitemi\hspace{\dimexpr\labelsep}Embedding Model $\rightarrow$ LLM.\\
\labelitemi\hspace{\dimexpr\labelsep}LLM $\rightarrow$ Guardrails.~\\
\labelitemi\hspace{\dimexpr\labelsep}RAG $\rightarrow$ augments LLM Input.\\
\labelitemi\hspace{\dimexpr\labelsep}LLM $\rightarrow$ augments subsequent Inputs, as context.} \\

\textbf{Output Details or Explanations.} & {\textit{LLM Generated Content}:\\
\labelitemi\hspace{\dimexpr\labelsep}RAG sources.~\\
\textit{Classifier Predictions}:\\
\labelitemi\hspace{\dimexpr\labelsep}Confidences~\\
\labelitemi\hspace{\dimexpr\labelsep}Top-5 classifications} & {\labelitemi\hspace{\dimexpr\labelsep}Inputs\\
\labelitemi\hspace{\dimexpr\labelsep}Outputs\\
\labelitemi\hspace{\dimexpr\labelsep}Weights, Activations} & {\labelitemi\hspace{\dimexpr\labelsep}Input to monitoring tools or services.\\
\labelitemi\hspace{\dimexpr\labelsep}Input to data collection processes; e.g. Active Learning.} \\

\textbf{Dataset.} & {\labelitemi\hspace{\dimexpr\labelsep}Raw Data, after encoding and transformation.\\
\labelitemi\hspace{\dimexpr\labelsep}Other Datasets.} & The training dataset is, to some extent, encoded into the model during training. Adversaries can, therefore,\textit{Indirectly Inspect}~the dataset by studying model behavior. & \labelitemi\hspace{\dimexpr\labelsep}Selections from RAG dataset serve as an LLM Input component. \\

\textbf{Model Parameters: Weights, Activations, Updates.} & {\labelitemi\hspace{\dimexpr\labelsep}Foundational Model\\
\labelitemi\hspace{\dimexpr\labelsep}Federated Learning Updates} & {\labelitemi\hspace{\dimexpr\labelsep}Dataset: Training.\\
\labelitemi\hspace{\dimexpr\labelsep}Validation Results.\\
\labelitemi\hspace{\dimexpr\labelsep}Updates influence weights and, therefore, future activations.\\
\labelitemi\hspace{\dimexpr\labelsep}Inputs influence activations} & \labelitemi\hspace{\dimexpr\labelsep}Input to Explainable AI (XAI) tools or agents. \\

\textbf{Validation or Monitoring Criteria.} & {\labelitemi\hspace{\dimexpr\labelsep}Corporate or Regulatory Governance\\
\labelitemi\hspace{\dimexpr\labelsep}cite MLCommons} & \labelitemi\hspace{\dimexpr\labelsep}Dataset: Validation & \labelitemi\hspace{\dimexpr\labelsep}Input to Validation agent (cite). \\

\textbf{Validation or Monitoring Results.} &
{\labelitemi\hspace{\dimexpr\labelsep}Statistics\\ \labelitemi\hspace{\dimexpr\labelsep}Recommended Actions}
& {
\labelitemi\hspace{\dimexpr\labelsep}Validation or Monitoring Criteria.\\
Monitoring Results also Influenced by:\\
\labelitemi\hspace{\dimexpr\labelsep}Inputs, Outputs, Output Details and Explanations.}
& \labelitemi\hspace{\dimexpr\labelsep}Output from Validation agent (cite). \\

\textbf{Hyperparameters: Model, Data, Training.} & {\labelitemi\hspace{\dimexpr\labelsep}Model architecture\\
\labelitemi\hspace{\dimexpr\labelsep}Model activation functions\\
\labelitemi\hspace{\dimexpr\labelsep}Data Encoding and Transformation configuration\\
\labelitemi\hspace{\dimexpr\labelsep}Data Selection criteria or metrics\\
\labelitemi\hspace{\dimexpr\labelsep}Training Learning Rate function\\
\labelitemi\hspace{\dimexpr\labelsep}Training Reward function} & \labelitemi\hspace{\dimexpr\labelsep}Validation and Monitoring Results & \labelitemi\hspace{\dimexpr\labelsep}Reinforcement learning for hyperparameter tuning cite \end{tblr}
\end{table*}
\begin{appendices}
\section{AI Asset Types Cheat-Sheet}
Table~\ref{tab:asset-types} provides an at-a-glance overview of the asset types introduced in Section~\ref{sec:asset_caps}.
It includes common components, dependencies, and relationships.
\end{appendices}

\end{document}